\documentclass{aa}
\usepackage{graphicx}
\usepackage{url}
\usepackage{mathrsfs,amssymb,amsmath}
\usepackage[usenames]{color}
\usepackage{subfigure}

\usepackage[varg]{txfonts}

\newcommand{\mW}{$\mathcal{W}$}
\newcommand{\mS}{$\mathcal{S}$}
\newcommand{\mP}{$\mathcal{P}$}
\newcommand{\zW}{$\zeta_{\mathcal{W}}$}
\newcommand{\zS}{$\zeta_{\mathcal{S}}$}

\newcommand{\bfl}{\begin{flushleft}}
\newcommand{\efl}{\end{flushleft}}

\newcommand{\ud}{\mathrm{d}} 
\newcommand{\be}{\begin{equation}}
\newcommand{\ee}{\end{equation}}

\newcommand{\zhh}{$\mathrm{\zeta}$}
\newcommand{\Ezhh}{\mathrm{\zeta}}

\begin{document}

\title{Protostars: Forges of cosmic rays?}

\author{M. Padovani\inst{1,2}, A. Marcowith\inst{1}, P. Hennebelle\inst{3}, and K. Ferri\`ere\inst{4}}

\authorrunning{M. Padovani et al.}

\institute{Laboratoire Univers et Particules de Montpellier, UMR 5299 du CNRS, Universit\'e de Montpellier, place E. Bataillon, cc072, 34095 Montpellier, France\\
\email{[Marco.Padovani,Alexandre.Marcowith]@umontpellier.fr}
\and
INAF--Osservatorio Astrofisico di Arcetri, Largo E. Fermi 5, 50125 
Firenze, Italy
\and 
CEA, IRFU, SAp, Centre de Saclay, 91191 Gif-Sur-Yvette, France\\
\email{patrick.hennebelle@cea.fr}
\and
IRAP, Universit\'e de Toulouse, CNRS, 9 avenue du Colonel Roche, BP 44346, F-31028 Toulouse Cedex 4, France\\
\email{katia.ferriere@irap.omp.eu}
}

\abstract 
{Galactic cosmic rays are particles presumably accelerated in supernova remnant shocks that propagate in the interstellar medium up to the densest parts of
molecular clouds, losing energy and their ionisation efficiency because of the presence of magnetic fields and collisions with molecular hydrogen. Recent observations hint at high levels of ionisation and at the presence of synchrotron emission in protostellar systems, 
which leads to an apparent contradiction.}
{We want to explain the origin of these cosmic rays accelerated within young protostars as suggested by observations.}
{Our modelling consists of a set of conditions that has to be satisfied in order to have an efficient cosmic-ray acceleration 
through diffusive shock acceleration. 
We analyse three main acceleration sites (shocks in accretion flows, along the jets, and on protostellar surfaces), then
we follow the propagation of these particles through the protostellar system up to the hot spot region.}
{We find that jet shocks can be strong accelerators of cosmic-ray protons, 
which can be boosted up to relativistic energies. Other promising acceleration sites are 
protostellar surfaces, where shocks caused by impacting material during the collapse phase are strong enough to accelerate cosmic-ray protons.
In contrast, accretion flow shocks are too weak to efficiently accelerate cosmic rays. Though cosmic-ray electrons are weakly accelerated, 
they can gain a strong boost to relativistic energies through re-acceleration in successive shocks.}
{We suggest a mechanism able to accelerate both cosmic-ray protons and electrons through the diffusive shock acceleration mechanism, 
which can be used to explain the high ionisation rate and the synchrotron emission observed towards protostellar sources.
The existence of an internal source of energetic particles can have a strong and unforeseen impact on the ionisation of the
protostellar disc, on the star 
and planet formation processes, and on the formation of pre-biotic molecules.}


\maketitle

\section{Introduction}
Cosmic rays (CRs), ordinary matter, and magnetic fields represent the fundamental elements of the Galaxy; they have comparable
pressures and are coupled together by electromagnetic forces (Ferri\`ere~\cite{f01}).
The interaction between CRs and the interstellar matter lays the foundation for the rich chemistry that is observed in molecular clouds.
In fact, as soon as the UV interstellar radiation field is absorbed, at about 4 magnitudes of visual extinction (McKee~\cite{m89}) and 
far from the X-ray flux produced by embedded protostars
(Krolik \& Kallman~\cite{kk83}; Silk \& Norman~\cite{sn83}), CRs are the main ionising agents 
of molecular hydrogen, the most abundant component of molecular clouds.
From this process, increasingly complex species are produced, allowing the characterisation of the physical and chemical properties of protostellar sources.

The key parameter in the calculation of molecular abundances from observations and from
chemical models is the CR ionisation rate, \zhh.
Determining the value of \zhh\ is not straightforward since the propagation of CRs inside a cloud  has to be accurately
described and a number of processes taken into account: energy losses (Padovani et al.~\cite{pgg09}),
magnetic field effects 
(Cesarsky \& V\"olk~\cite{cv78}; Chandran~\cite{c00}; 
Padovani \& Galli~\cite{pg11}; Padovani \& Galli~\cite{pg13}; Padovani et al.~\cite{phg13}), 
and screening due to self-generated Alfv\'en waves in the plasma
(Skilling \& Strong~\cite{ss76}; Cesarsky \& V\"olk~\cite{cv78}; Hartquist et al.~\cite{hd78}; Rimmer et al.~\cite{rh12}; Morlino \& Gabici~\cite{mg15}).

In addition to the chemistry, CRs play another important role in regulating the formation of protostellar discs. 
A magnetic field entrained in a collapsing cloud  brakes 
the rotational motions as long as the field is frozen in the gas  
(see e.g. Galli et al.~\cite{gl06}; Mellon \& Li~\cite{ml08}; Hennebelle \& Fromang~\cite{hf08}).
One of the speculated mechanisms that mitigates the magnetic braking effect relies on non-ideal
magnetohydrodynamic effects, namely ambipolar diffusion, Hall, and Ohmic diffusion 
(Shu et al.~\cite{sg06}; Dapp \& Basu~\cite{db10}; Krasnopolsky et
al.~\cite{kl11}; Braiding \& Wardle~\cite{bw12a,bw12b}; Masson et al.~\cite{mc15}; Tomida et al.~\cite{to15}).
The associated diffusion coefficients depend on the abundance of the charged species, which in turn is predicted by the CR ionisation rate.   
Padovani et al.~(\cite{phg13,pg14}) showed that, at least in the formation process of low-mass protostellar discs, 
a proper treatment of CR propagation can lead to very low values of \zhh. As a consequence,
in the central region of a collapsing cloud
the coupling between gas and magnetic field is weaker than usually assumed, flux freezing is no longer valid, 
and the influence of the magnetic field on the collapse is reduced. 

Cleeves et al.~(\cite{ca13}) studied the inhibition of CR propagation in protoplanetary discs of Class~II protostars as a result of magnetised stellar
winds. They found that, in addition to X-ray ionisation from the central star, \zhh\ is set by short-lived radionuclides and it is of the 
order of $10^{-19}$~s$^{-1}$.
On the other hand, Ceccarelli et al.~(\cite{cd14}) and Podio et al.~(\cite{pl14}) found very high values of \zhh\ towards
two protostars (OMC-2 FIR~4 and L1157-B1). 
These high values of \zhh\ cannot be due to the interstellar CR flux since the column density is too high, which damps the propagation of
interstellar CRs (Padovani et al.~\cite{phg13}). Following the same reasoning, the interstellar electron flux cannot explain
the synchrotron emission observed towards the bow shock of DG~Tau by Ainsworth et al.~(\cite{as14}).

The purpose of this paper is to investigate the possibility of accelerating particles, i.e.  of generating local CRs, inside
or in the immediate vicinity of a protostar. Our investigation is justified by simple arguments on the energetics of the system.
The gravitational luminosity of an accretion shock on the surface of a protostar reads
\be\label{Lgrav}
L_{\rm grav}=\frac{GM\dot{M}}{R_{\rm sh}}\,,
\ee
where $G$ is the gravitational constant, $M$ is the protostellar mass, $\dot{M}$ is the accretion rate,
and $R_{\rm sh}$ is the shock radius. If we consider the gravitational collapse of an early Class 0 protostar with $M=0.1~M_{\odot}$,
$\dot{M}=10^{-5}~M_{\odot}$~yr$^{-1}$ (e.g. Shu et al.~\cite{sa87}; Belloche et al.~\cite{ba02}), $R_{\rm sh}=2\times10^{-2}$~AU (Masunaga \& Inutsuka~\cite{mi00}), then
$L_{\rm grav}=3\times10^{34}$~erg~s$^{-1}$. The luminosity of the interstellar CRs impinging on a molecular cloud core can be estimated by
\be\label{Lcr}
L_{\rm CR}=R_{\rm core}^{2}V_{\rm A}\epsilon_{\rm CR}\,,
\ee
where $R_{\rm core}$ is the core radius; $V_{\rm A}$ is the Alfv\'en speed in the surrounding medium, supposed to be the warm neutral medium;
and $\epsilon_{\rm CR}$ is the energy density of the interstellar CRs based on the latest 
Voyager~1 observations (Stone et al.~\cite{sc13}; Ivlev et al.~\cite{ip15}). Here we adopt $R_{\rm core}=0.1$~pc, \mbox{$V_{\rm A}=9.3\times10^{5}$~cm~s$^{-1}$} (based on $n_{\rm H}=0.5$~cm$^{-3}$ and $B=3~\mu$G; Ferri\`ere~\cite{f01}), 
and $\epsilon_{\rm CR}=1.3\times10^{-12}$~erg~cm$^{-3}$, 
then $L_{\rm CR}=1.2\times10^{29}$~erg~s$^{-1}\ll L_{\rm grav}$. In previous studies we demonstrated that the interstellar CR flux is strongly
attenuated at high column densities (Padovani et al.~\cite{pgg09}, \cite{phg13}), and so we
expect $\epsilon_{\rm CR}$ at the protostellar surface to be much lower than its interstellar value and, a fortiori, 
$L_{\rm CR}\lll L_{\rm grav}$ close to the protostar. Thus, 
if a small fraction of the gravitational energy can be used to produce local CRs, they could easily be dominant over the ISM ones.
For a massive star the gravitational energy available to generate high-energy CRs is even higher since $\dot{M}$ could be
as high as $10^{-3}~M_{\odot}$~yr$^{-1}$ and in principle their $\gamma$ emission can be observed
(see e.g. Araudo et al.~\cite{ar07}; Bosch-Ramon et al.~\cite{br10}; Munar-Adrover et al.~\cite{map11}).

The organisation of the paper is the following.
In Sect.~\ref{Sect2} we examine the acceleration processes that can take place in protostellar shocks, carefully analysing the conditions leading to
particle acceleration (see also Padovani et al.~\cite{phm15}), and
in Sect.~\ref{PCR} we evaluate the pressure of accelerated CRs.
In Sect.~\ref{fulfilment} we verify in what part of the protostar these conditions are fulfilled, 
and in Sect.~\ref{spectra} we evaluate the maximum energy that can be reached
and the emerging spectrum at the shock surface. In Sect.~\ref{energylossregime} we describe
the propagation mechanism of local CRs, how they can be re-accelerated at the reverse bow shock of a jet, and 
how they propagate in the hot spot region.
In Sect.~\ref{ionrates} we derive the CR ionisation rate along the jet and the temperature profile of the protostellar disc.
In Sect.~\ref{observations}
we use our modelling to explain observational results, and in Sect.~\ref{conclusions} we summarise our conclusions.
Finally, in the appendices we give more details 
on alternative acceleration sites and mechanisms (Appendix~\ref{app:othermechanisms}),
the damping of turbulence in a jet (Appendix~\ref{app:dampingrate}),
the collisional character of the shocks and the thermal equilibration (Appendix~\ref{app:collisionless}),
the calculation of the ion-neutral damping condition (Appendix~\ref{app:Ecut}),
and the relevance of using a steady-state model (Appendix~\ref{app:lifetime}).

\section{Cosmic-ray acceleration in protostellar shocks}\label{Sect2}

Protostars are classified as a function of their spectral energy distribution in the near- and mid-infrared domain (Adams et al.~\cite{al87}, Andr\'e
et al.~\cite{am94}). 
These protostars are surrounded by dusty envelopes that absorb and re-emit at infrared wavelengths the energy irradiated by the central forming star,
and this envelope becomes  increasingly  optically thin during the evolutionary sequence. 
The most embedded objects are named Class 0; they have a very weak emission at infrared wavelengths with relevant emission in the 
submillimetre spectral range. At this stage the envelope has already started its gravitational collapse and collimated polar outflows and jets 
can be observed, but  the envelope mass is still much larger than the central condensation. 
Class I objects represent the following step in the evolution of prestellar cores where
the envelope  progressively dissolves  through
accretion processes onto the central object together with ejecta in the form of outflows and jets.
These objects are detectable in the infrared, but  not at optical wavelengths.
Finally, Class II and Class III objects are pre-main sequence stars,
which are the most evolved stages of a protostar. These objects can be observed at both optical and 
infrared wavelengths. Class II sources present an optically thick circumstellar disc, while in Class III the 
disc is optically thin, but both are lacking a circumstellar envelope.

In the following we identify a number of possible particle acceleration sites in protostars. 
Some of them are peculiar in the first stages of protostar collapse (mainly Class 0), such as accretion flows 
in the protostellar envelope; others, such as jets, are common to Class~0 and the more evolved protostar classes.
This work focuses on shock acceleration, but alternative processes are discussed in Appendix~\ref{app:othermechanisms}.

\subsection{Diffusive shock acceleration}\label{conditionsDSA}
Also known as first-order Fermi acceleration, diffusive shock acceleration (DSA) is a process where charged particles systematically gain energy while crossing a shock front. 
Multiple shock crossings allow the particle energy to rapidly increase and to  reach the relativistic domain. The motion of particles back and 
forth from upstream to downstream requires the presence of magnetic fluctuations that produce a scattering of the pitch angle, which is the angle 
between the particle's velocity and the mean magnetic field. As a consequence, the distribution in momentum space 
of the emerging accelerated particles is set by the ratio of the relative energy gain in an acceleration (or Fermi) cycle to the probability of 
being advected with the scattering centres downstream. 
In the limit of strong shocks, when the magnetic field is not dynamically dominant, 
this ratio only depends on the shock compression ratio $r$ 
(see Eq.~\ref{compressionratio}), and once $r$ is fixed the shock distribution is a power law. 
This process is described in several reviews (e.g. Drury~\cite{d83}; Kirk~\cite{k94}).
In the following subsections we describe in greater
detail the conditions (summarised in Padovani et al.~\cite{phm15}) that have to be satisfied in order to effectively accelerate particles through the DSA mechanism: 
criteria that need to be fulfilled regardless of the origin of the magnetic disturbances causing the particle scatterings 
(Sects.~\ref{csva} and~\ref{colllosses}), those associated with the shock age and geometry 
(Sect.~\ref{agegeometry}), and with magnetic field fluctuations that are self-generated by the particles 
themselves (Sects.~\ref{ionneutralfriction} and~\ref{PCR}).

All the conditions limiting the maximum energy of the accelerated particles are written as functions of the upstream
flow velocity in the shock 
reference frame
\be\label{velref}
U=v_{\rm fl}-v_{\rm sh}\,,
\ee
where $v_{\rm fl}$ and $v_{\rm sh}$ are the flow and shock velocities, respectively, both taken in the observer reference frame.
We note  that $v_{\rm fl}$ will later be equated to $v_{\rm acc}$ or $v_{\rm jet}$ depending on the acceleration site.

\begin{figure}[!htb]
\begin{center}
\resizebox{\hsize}{!}{\includegraphics{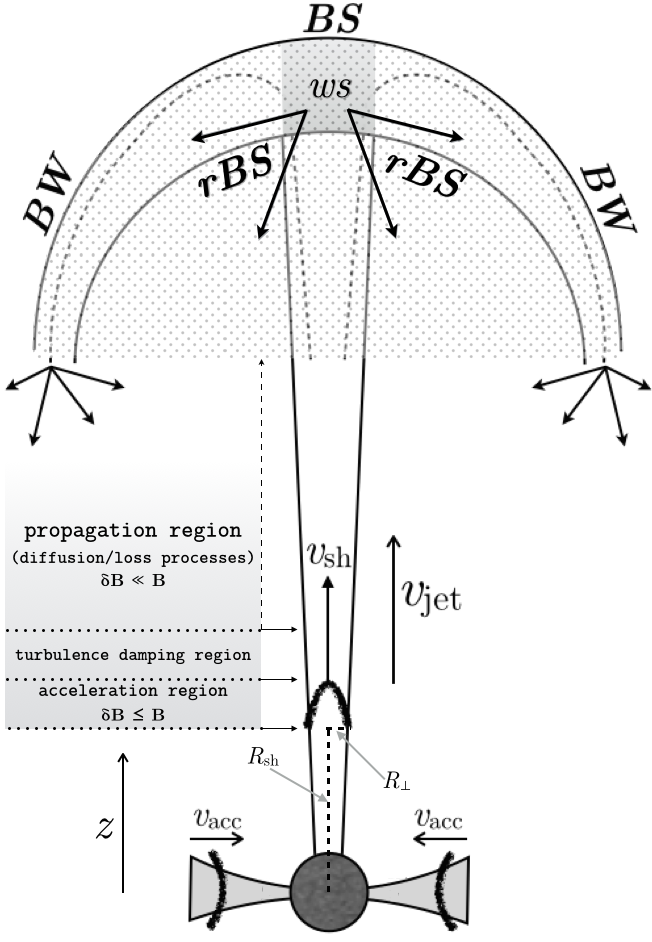}}
\caption{Sketch of the protostar configuration employed in the paper. Accretion flow, jet, and shock velocities 
($v_{\rm acc}$, $v_{\rm jet}$, and $v_{\rm sh}$, respectively) are in the observer reference frame.
Shock and transverse radii are labelled  $R_{\rm sh}$ and $R_{\perp}$, respectively.
Bow shock, reverse bow shock, bow wings, and working surface are labelled  {\em BS}, {\em rBS}, {\em BW}, and {\em ws}, respectively.
The shaded areas show the regions where CR acceleration takes place (Sect.~\ref{conditionsDSA}) and where 
turbulence is damped and
particle propagation occurs (Sect.~\ref{energylossregime}). The dot-filled region corresponds to the hot spot region
(Sect.~\ref{twozonemodel}).}
\label{sketch}
\end{center}
\end{figure}

Figure~\ref{sketch} outlines the configuration of a protostar used for our modelling. The shock in the accretion
flow is assumed to be stationary ($v_{\rm sh}=0$). 
Shocks inside jets  move more slowly than the flow (see Sect.~\ref{jet}) so that the upstream region is close to the protostar. 
The reverse bow shock (also known as Mach disc) 
and the bow shock, when observed,
 usually
move slowly  or are stationary as well (Caratti o Garatti \& Eisl\"offel~\cite{ce09}).
After passing through the jet shock, the gas flow spreads out until it yields a bow shock with its reverse bow shock.
In the intershock region, known as the working surface, the pressure gradient forces perpendicular to the jet are so high that the gas is 
ejected radially and  propagates into the surrounding region, called the hot spot region (e.g. Stahler \& Palla~\cite{sp05}), 
through the bow wings.
Multiple shocks are observed, and so throughout the paper we only account  for the presence of a single jet shock. 

We briefly
discuss the consequences of recurring acceleration processes in Sect.~\ref{conclusions}, and
in Appendix~\ref{app:othermechanisms} we describe  alternative mechanisms that may provide 
efficient CR acceleration in addition to DSA.

\subsection{Condition on shock velocity}\label{csva}
In order to have efficient particle acceleration, the flow has to be supersonic and super-Alfv\'enic. These two conditions are combined into the 
 relation
\begin{eqnarray}\label{Uhighercsva}
\frac{U}{10^{2}~\mathrm{km~s^{-1}}}&>&\max\left\{9.1\times10^{-2}\left[\gamma_{\rm ad} (1+x)%
\left(\frac{T}{10^{4}~\mathrm{K}}\right)\right]^{0.5},\right.\\\nonumber 
&&\left.2.2\times10^{-4}\left(\frac{n_{\rm H}}{10^{6}~\mathrm{cm^{-3}}}\right)^{-0.5}\left(\frac{B}{10~\mathrm{\mu G}}\right)\right\}\,,
\end{eqnarray}
where
$\gamma_{\rm ad}$ is the adiabatic index, $T$ the upstream temperature, $n_{\rm H}$ the total number density of hydrogen, 
$x$ the ionisation fraction,
and $B$ the magnetic field strength. 
The two terms inside the braces on the right-hand side of Eq.~(\ref{Uhighercsva}) are the ambient (or upstream) sound speed, $c_{s}$, and
Alfv\'en speed, $V_{\rm A}$, respectively, of the total gas, i.e. when ions and neutrals are coupled, in units of $10^{2}$~km~s$^{-1}$.
The contribution of helium and heavier species in the background plasma is neglected.

The electron temperature, $T_{e}$, 
can be estimated by observations, and downstream of a shock any temperature difference 
is rapidly thermalised so that the proton temperature, $T_{p}$, can be safely assumed to be equal to $T_{e}$.
In fact, even if in the presence of collisionless shocks, such that
the passage of the shock creates a gradient between the temperatures of the two species 
($m_{p}T_{e}\simeq m_{e}T_{p}$), it is possible to demonstrate that the time needed to reach temperature equilibrium 
is shorter than the time between two consecutive shocks (see Appendix~\ref{app:collisionless}).

\subsection{Condition on low-energy cosmic-ray acceleration: collisional losses}\label{colllosses}
We are interested in the acceleration of low-energy ($\lesssim100~\mathrm{MeV}-1~\mathrm{GeV}$) CRs 
since they are responsible for the bulk of the ionisation. 
We have
to verify that the shock acceleration rate is higher than the collisional loss rate, 
$t^{-1}_{\rm acc}>t^{-1}_{\rm loss}$.
Adapting
Drury et al.~(\cite{dd96}) to account for both parallel and perpendicular shock configurations,\footnote{A parallel/perpendicular shock is when the shock normal is parallel/perpendicular 
to the ambient magnetic field. More complex oblique geometries are not considered.} the acceleration rate is given by
\begin{eqnarray}\label{nuacc}
t^{-1}_{\rm acc}&=&\frac{1.1\times10^{-8}}{\gamma-1}\frac{k_{\rm u}^{\alpha}(r-1)}{r[1+r(k_{\rm d}/k_{\rm u})^{\alpha}]}%
\tilde\mu^{-1}\\\nonumber
&&\times\left(\frac{U}{\mathrm{10^{2}~km~s^{-1}}}\right)^{2}%
\left(\frac{B}{\mathrm{10~\mu G}}\right)~\mathrm{s^{-1}}\,,
\end{eqnarray}
where $\tilde\mu=m/m_{p}$ is the particle mass normalised to the proton mass, $k_{\rm u}$ and $k_{\rm d}$ are the diffusion coefficient 
in the upstream and downstream media, respectively, defined by
\be \label{diffusioncoeff}
k_{\rm u}=\left(\frac{\kappa_{\rm u}}{\kappa_{\rm B}}\right)^{-\alpha}=%
\left(\frac{3eB}{\gamma\beta^{2}m_{p}c^{3}}\kappa_{\rm u}\right)^{-\alpha}\,,
\ee
with $\kappa_{\rm B}$ the Bohm diffusion coefficient, $e$ the elementary charge, 
$\gamma$ the Lorentz factor, and $\beta=v/c=\gamma^{-1}\sqrt{\gamma^{2}-1}$ ($v$ and $c$ are the speed of the particle and the speed of light,
respectively),
and they can possibly be dependent on the particle momentum (Jokipii~\cite{j87}).
For a parallel shock $\alpha=-1$ and $k_{\rm u}=k_{\rm d}$, while for a perpendicular shock $\alpha=1$ and,
since the magnetic field is compressed by a factor $r$, 
$k_{\rm u}=rk_{\rm d}$ (Priest~\cite{p94}).
The shock compression ratio, $r$, is given by
\be\label{compressionratio}
r=\frac{\left(\gamma_{\rm ad}+1\right)M_{s}^{2}}{\left(\gamma_{\rm ad}-1\right)M_{s}^{2}+2}\,,
\ee
where
\be\label{Mson}
M_{s}=\frac{U}{c_{s}}
\ee
is the sonic Mach number and in the following the adiabatic index is set to $\gamma_{\rm ad}=5/3$. 
Obviously, Eq.~(\ref{compressionratio}) supposes that the shock is adiabatic, an assumption that might not always be satisfied. 
Owing to the dynamical nature of protostars, shocks might sometimes be radiative, in which case 
the DSA scenario should be modified as, for instance, injection from the thermal pool might not work.
In this case, alternative acceleration mechanisms should be 
considered (see Appendix~\ref{app:othermechanisms}). In addition, Eq.~(\ref{compressionratio}) is obtained in hydrodynamics, 
whereas DSA requires the presence of a magnetic field. Here we verified that magnetic pressure on each side of the shock front 
is much lower than the ram pressure of the flow so that the magnetic field is dynamically negligible and 
Eq.~(\ref{compressionratio}) remains valid.
However, the magnetic field is 
relevant for particle scattering. 

The collisional energy loss rate is given by
\begin{equation} \label{nuloss}
t^{-1}_{\rm loss}=3.2\times10^{-9}\frac{\beta}{\gamma-1}\tilde\mu^{-1}%
\left(\frac{n_{\rm H}}{\mathrm{10^{6}~cm^{-3}}}\right)%
\left[\frac{L(E)}{\mathrm{10^{-25}~GeV~cm^{2}}}\right]~\mathrm{s^{-1}}\,,
\end{equation}
where
$L(E)$ is the energy loss function (Padovani et al.~\cite{pgg09}), which was
extended to lower energies to include Coulomb losses. These functions 
are given for protons by Mannheim \& Schlickeiser~(\cite{ms94}),
\be
\frac{L_{C,p}(E)}{\mathrm{10^{-25}~GeV~cm^{2}}}=0.1\frac{x\beta}{\beta_{\rm th}^{3}+\beta^{3}}\,,
\ee
where $\beta_{\rm th}=2\times10^{-3}(T/\mathrm{10^{4}~K})^{0.5}$. 
For electron Coulomb losses, we use the analytic fit by Swartz et al.~(\cite{sn71}),
\begin{equation}
\frac{L_{C,e}(E)}{\mathrm{10^{-25}~GeV~cm^{2}}}=7.3\times10^{-3}\frac{x}{\beta}\left(\frac{E}{1~\mathrm{GeV}}\right)^{-0.44}%
\left(\frac{E-E_{\rm th}}{E-0.53E_{\rm th}}\right)^{2.36}\,,
\end{equation}
where $E_{\rm th}$ is the electron thermal energy.
Synchrotron losses, $L_{S,e}$, were included in the energy loss function 
for electrons and they read (Schlickeiser~\cite{s02})
\begin{equation}\label{synlosses}
\frac{L_{S,e}(E)}{10^{-25}~\mathrm{GeV~cm^{2}}}=2\times10^{-14}\frac{\gamma^{2}}{\beta}%
\left(\frac{n_{\rm H}}{10^{6}~\mathrm{cm^{-3}}}\right)^{-1}\left(\frac{B}{10~\mu\mathrm{G}}\right)^{2}\,.
\end{equation}
The maximum energy of accelerated particles set by energy losses, $E_{\rm loss}$, is found when $t_{\rm acc}^{-1}=t_{\rm loss}^{-1}$,
leading to
\begin{eqnarray}\label{FEloss}
\beta\left[\frac{L(E)}{\mathrm{10^{-25}~GeV~cm^{2}}}\right]&=&3.4\frac{k_{\rm u}^{\alpha}(r-1)}{r[1+r(k_{\rm d}/k_{\rm u})^{\alpha}]}%
\\\nonumber
&&\times\left(\frac{U}{10^{2}~\mathrm{km~s^{-1}}}\right)^{2}%
\left(\frac{n_{\rm H}}{10^{6}~\mathrm{cm^{-3}}}\right)^{-1}%
\left(\frac{B}{10~\mu\mathrm{G}}\right)\,.
\end{eqnarray}

\subsection{Condition on CR acceleration: ion-neutral friction}\label{ionneutralfriction}

In this work we assume that CR scattering occurs in the so-called quasi-linear regime. This assumes that the level of magnetic 
fluctuations produced by the CRs themselves or by the background turbulence is lower than the background large-scale magnetic 
field or at most equal to it. In that regime the CR's pitch angle is only slightly deflected during an interaction with a magnetic 
perturbation so that the gyromotion around the background magnetic field can be retained as a good approximation of the trajectory
(Schlickeiser~\cite{s02}).

The main limit on the possibility of CR acceleration is given by the presence
of an incompletely ionised medium. The collision rate between ions and neutrals
can actually be  high enough to decrease the effectiveness of DSA.
The presence of neutrals damps the CR self-generated fluctuations that allow the CRs to move back and forth across the shock multiple times.
Ions and neutrals are effectively decoupled if the wave frequency is higher than the ion-neutral collision frequency, otherwise neutrals 
take part in the coherent oscillations between ions and Alfv\'en waves. Here we consider resonant waves whose pulsation satisfies the condition
$\omega\sim V_{\rm A}/r_g$, where the gyroradius is $r_g\sim\gamma \beta \tilde\mu m_p c^2/(e B)$. 
Following Eq.~(11) in Drury et al.~(\cite{dd96}) and accounting for the fact that CRs are not fully relativistic, we find that the critical
energy separating these two regimes, $E_{\rm coup}$, is derived by solving the following relation:
\be\label{Ecoup}
\gamma\beta=8.5\times10^{-7}\tilde\mu^{-1}\left(\frac{T}{10^{4}~\mathrm{K}}\right)^{-0.4}%
\left(\frac{n_{\rm H}x}{10^{6}~\mathrm{cm^{-3}}}\right)^{-1.5}\left(\frac{B}{10~\mathrm{\mu G}}\right)^{2}\,.
\ee
If the CR energy is higher than $E_{\rm coup}$, ions and neutrals are coupled.

The upper cut-off energy due to wave damping, $E_{\rm damp}$, is set by requiring that the flux of locally accelerated CRs
advected downstream by the flow be equal to the flux of CRs lost upstream because of the lack of waves (due to wave
damping) to confine the CRs. Following Drury et al.~(\cite{dd96}),
using their exact equation for the wave damping rate,
accounting for departures from fully relativistic behaviour, and assuming $U$ to be much higher than the Alfv\'en speed,
$E_{\rm damp}$ follows from\footnote{See Appendix~\ref{app:Ecut} for details on the
derivation of $E_{\rm damp}$.}
\begin{eqnarray}\label{Edamp}
\gamma\beta^{2}&=&8.8\times10^{-5}\tilde\mu^{-1}\Xi(1-x)^{-1}%
\left(\frac{U}{10^{2}~\mathrm{km~s^{-1}}}\right)^{3}%
\left(\frac{T}{10^{4}~\mathrm{K}}\right)^{-0.4}\\\nonumber%
&&\times\left(\frac{n_{\rm H}}{10^{6}~\mathrm{cm^{-3}}}\right)^{-0.5}
\left(\frac{B}{10~\mathrm{\mu G}}\right)^{-4}%
\left(\frac{\widetilde{P}_{\rm CR}}{10^{-2}}\right)\,,
\end{eqnarray}
where
\be\label{XI}
\Xi=\left(\frac{B}{10~\mathrm{\mu G}}\right)^{4}+1.4\times10^{12}\tilde\mu^{2}\gamma^{2}\beta^{2}x^{2}%
\left(\frac{T}{10^{4}~\mathrm{K}}\right)^{0.8}%
\left(\frac{n_{\rm H}}{10^{6}~\mathrm{cm^{-3}}}\right)^{3}
,\ee
and
\be\label{Pcrprime}
\widetilde{P}_{\rm CR}=\frac{P_{\rm CR}}{n_{\rm H}m_{p}U^{2}}
\ee
is the fraction of the shock ram pressure going into CR acceleration (see also Sect.~\ref{PCR}).

The normalisation is taken with respect to the proton mass since the contribution of electrons to the total CR pressure is negligible.
Both Eq.~(\ref{Ecoup}) and Eq.~(\ref{Edamp}) are valid for $T\in[10^{2},10^{5}]$~K.

If $E_{\rm damp}>E_{\rm coup}$, then $E_{\rm damp}$ is in the coupled regime, i.e. neutrals coherently move with ions and
ion-generated waves are weakly damped. The condition $E_{\rm damp}>E_{\rm coup}$ can be written by combining Eqs.~(\ref{Ecoup}) and~(\ref{Edamp}) as
\begin{eqnarray}\label{Rratio}
\mathscr{R}&=&\frac{10^{2}}{\beta}\Xi%
~x^{1.5}%
\left(\frac{U}{10^{2}~\mathrm{km~s^{-1}}}\right)^{3}\\\nonumber%
&&\times\left(\frac{n_{\rm H}}{10^{6}~\mathrm{cm^{-3}}}\right)
\left(\frac{B}{10~\mathrm{\mu G}}\right)^{-6}%
\left(\frac{\widetilde{P}_{\rm CR}}{10^{-2}}\right)>1\,.
\end{eqnarray}
In the following sections we consider shocks in three types of environments: in jets, in accretion flows in the collapsing envelopes,
and on the surfaces of protostars.
Using the range of parameters in Table~\ref{paramspace}, we find $\mathscr{R}\ll1$ in protostellar envelopes 
(see Sect.~\ref{accflows}).
Therefore, the following two conditions on shock age and geometry (Sect.~\ref{agegeometry})
are uniquely discussed with reference to shocks in jets and on protostellar surfaces.

\subsection{Conditions arising from shock age and geometry}\label{agegeometry}
The limit on the maximum energy determined by the age of the shock, $E_{\rm age}$, is found when the acceleration time, given by the inverse of Eq.~(\ref{nuacc}), is equal to 
the age of the shock, $t_{\rm age}$. The latter can be assumed to be of the order of the dynamical time of the jet
($\gtrsim10^{3}$~yr, de Gouveia Dal Pino~\cite{dg95})
or equal to the accretion time in the case of a surface shock,
i.e. the time needed for a mass element in the envelope to reach the central protostar
($\sim10^{5}$~yr, Masunaga \& Inutsuka~\cite{mi00}). 
Then, $E_{\rm age}$ is computed from
\begin{eqnarray}\label{Eage}
\gamma-1&=&3.2\times10^{2}\frac{k_{\rm u}^{\alpha}(r-1)}{r[1+r(k_{\rm d}/k_{\rm u})^{\alpha}]}\tilde\mu^{-1}\\\nonumber%
&&\times\left(\frac{U}{10^{2}~\mathrm{km~s^{-1}}}\right)^{2}%
\left(\frac{B}{10~\mathrm{\mu G}}\right)%
\left(\frac{t_{\rm age}}{10^{3}~\mathrm{yr}}\right)\,.
\end{eqnarray}

When $\mathscr{R}>1$, the most stringent constraint is given by the geometry of the shock. In particular, the upstream diffusion length, 
$\lambda_{\rm u}=\kappa_{\rm u}/U$, has to be at most a given fraction $\epsilon<1$ of the shock radius,
namely the distance from the source, $R_{\rm sh}$. 
For a Class~0 protostar, we assume the accretion on the protostellar surface to be still spherical and not driven by accretion
columns from the inner disc. 
We also assume the shock to be planar since the CR's mean free path around the
shock is smaller than the transverse size of the jet, $R_{\perp}$.
Because of the spherical geometry,
no transverse escape either upstream or downstream is expected.
In contrast, in the jet configuration 
CRs may also escape in the transverse direction.
The maximum energy due to upstream escape losses, $E_{\rm esc,u}$, follows from 
\be\label{FEescu}
\gamma\beta^{2}=4.8\mathscr{M}k_{\rm u}^{\alpha}\tilde\mu^{-1}%
\left(\frac{U}{10^{2}~\mathrm{km~s^{-1}}}\right)%
\left(\frac{B}{10~\mathrm{\mu G}}\right)\,,
\ee
where 
\be
\mathscr{M}=\frac{\epsilon R_{\rm sh}}{10^{2}~\mathrm{AU}}
\ee
or
\be
\mathscr{M}=\min\left[\frac{\epsilon R_{\rm sh}}{10^{2}~\mathrm{AU}}, \frac{R_{\perp}}{10^{2}~\mathrm{AU}}\right] 
\ee
for a shock
on the protostellar surface or in the jet, respectively.
In the following we assume $\epsilon=0.1$ (Berezhko et al.~\cite{be96}).

Since jet shocks have a small transverse dimension, there is a further condition for the escape of CRs downstream:
the maximum energy due to downstream escape losses, $E_{\rm esc,d}$, is found when
the acceleration time, the inverse of Eq.~(\ref{nuacc}), is equal to the downstream 
diffusion time, $t_{\rm diff,d}$, which is given by\footnote{The factor 4 in the denominator of Eq.~(\ref{tdiffd}) comes from the fact that the diffusion process in the perpendicular direction is in two dimensions.}
\be\label{tdiffd}
t_{\rm diff,d}=\frac{R_{\perp}^{2}}{4\kappa_{\rm d}}\,.
\ee
Then, $E_{\rm esc,d}$ follows from
\begin{eqnarray}\label{FEescd}
\gamma\beta^{2}(\gamma-1)&=&2.1\mathscr{C}\frac{(k_{\rm u}k_{\rm d})^{\alpha}(r-1)}{r[1+r(k_{\rm d}/k_{\rm u})^{\alpha}]}\tilde\mu^{-1}\\\nonumber%
&&\times\left(\frac{U}{10^{2}~\mathrm{km~s^{-1}}}\right)^{2}%
\left(\frac{B}{10~\mathrm{\mu G}}\right)^{2}%
\left(\frac{R_{\perp}}{10^{2}~\mathrm{AU}}\right)^{2}\,,
\end{eqnarray}
with $\mathscr{C}=1$ or $r^{2}$ for a parallel or a perpendicular shock, respectively.

Finally, if the shock is supersonic and super-Alfv\'enic (Eq.~\ref{Uhighercsva}) and if $\mathscr{R}>1$ (Eq.~\ref{Rratio}),
the maximum energy reached by a particle is 
\be\label{Emax}
E_{\rm max}=\min[E_{\rm loss},E_{\rm damp},E_{\rm age},E_{\rm esc,u},E_{\rm esc,d}]\,,
\ee
where $E_{\rm loss}$, $E_{\rm damp}$, $E_{\rm age}$, $E_{\rm esc,u}$, and $E_{\rm esc,d}$ are given by
Eqs.~(\ref{FEloss}), (\ref{Edamp}), (\ref{Eage}), (\ref{FEescu}), and (\ref{FEescd}), respectively.

\section{Pressure of accelerated CRs}\label{PCR}
The value of the maximum energy is constrained by the fraction of the shock ram pressure that is channelled to CR pressure
(see Eqs.~\ref{Edamp} and \ref{Pcrprime}).
We predict both non-relativistic and mildly relativistic CRs and we checked a posteriori that there is no
strong back-reaction. This means that the upstream medium is not warned by these CRs that a shock is coming and
we can safely assume that the shock and the DSA process are unmodified.
For this reason, we can describe the CR momentum distribution function at the shock surface 
in the test-particle regime with a power law of momentum
\be\label{fp}
f(p)=f_{0}\left(\frac{p}{p_{\rm inj}}\right)^{-q}\,,
\ee
with $f_{0}$ normalisation constant;  $p_{\rm inj}<p<p_{\rm max}$,
where $p_{\rm inj}$ is the injection momentum (Eq.~\ref{injmom}) and $p_{\rm max}$ is the maximum momentum (given by
Eq.~\ref{Emax}); 
and $q=3r/(r-1)$
is the CR distribution index in the test-particle limit, with $r$ the compression ratio at the shock 
surface (Eq.~\ref{compressionratio}).
By definition, $f(p)$ is related to the CR density per unit volume, $n_{\rm CR}$, by
\be\label{nCR}
n_{\rm CR}=4\pi\int p^{2}f(p)\ud p=4\pi{\mathscr I}_{1}p_{\rm inj}^{3}f_{0}\,,
\ee
where 
\be\label{I1}
{\mathscr I}_{1}=\frac{1}{q-3}\left[1-\left(\frac{p_{\rm max}}{p_{\rm inj}}\right)^{3-q}\right]\,.
\ee
Assuming an efficient pitch-angle scattering and hence an isotropic CR distribution, the CR pressure reads
\be\label{partpress}
P_{\rm CR}=\frac{4\pi}{3}\int p^{3}vf(p)\ud p=\frac{4\pi}{3}\mathscr{I}_{2}(m_{p}c)^{4}cf_{0}\left(\frac{p_{\rm inj}}{m_{p}c}\right)^q\,,
\ee
where $v$ is the CR velocity with 
\be\label{I2}
{\mathscr I}_{2}=\int_{p_{\rm inj}/m_{p}c}^{p_{\rm max}/m_{p}c}\frac{\widetilde{p}^{\,4-q}}{\sqrt{\widetilde{p}^{\,2}+1}}\ud\widetilde{p}\,,
\ee
where $\widetilde{p}=p/(m_{p}c)$\,. 

Berezhko \& Ellison~(\cite{be99}) give the expressions for the normalised CR pressure
in non-relativistic and relativistic regimes 
and also in the transition region. 
Eliminating $f_{0}$ by equating Eqs.~(\ref{nCR}) and (\ref{partpress}), the sum of these pressures gives
\be\label{eqPcr}
\widetilde{P}_{\rm CR}=\eta r\left(\frac{c}{U}\right)^{2}\widetilde{p}_{\rm inj}^{\,a}%
\left(\frac{1-\widetilde{p}_{\rm inj}^{\,b_{1}}}{2r-5}+\frac{\widetilde{p}_{\rm max}^{\,b_{2}}-1}{r-4}\right)\,,
\ee
where 
$a=3/(r-1)$, $b_{1}=(2r-5)/(r-1)$, and $b_{2}=(r-4)/(r-1)$.
The parameter $\eta\in[10^{-6},10^{-3}]$ (Berezhko \& Ellison~\cite{be99})
is the shock efficiency, which represents the fraction of particles extracted from the thermal plasma
and injected into the acceleration process by a shock.
In the context of supernova remnants at relativistic energies, it is assumed that at least 10\% of the shock ram pressure goes into
CR acceleration (Berezhko \& Ellison~\cite{be99}). 
In contrast, protostellar shocks are expected to be much less energetic events, with $\widetilde{P}_{\rm CR}\ll10\%$;
accordingly, in the following we assume $\eta\in[10^{-6},10^{-5}]$.
Following Blasi et al.~(\cite{bg05}), the minimum (or injection) momentum, $p_{\rm inj}$, of a particle able to cross the shock that enters
the acceleration process is related to the thermal particle momentum, $p_{\rm th}$, through
\be\label{injmom}
p_{\rm inj}=\lambda p_{\rm th} = \lambda m c_{s,\rm d}\,,
\ee
where $c_{s,\rm d}$ is the sound speed in the downstream region in the strong shock limit given by (Berezhko \& Ellison~\cite{be99})
\be
c_{s,\rm d}=\frac{U}{r}\sqrt{\gamma_{\rm ad}(r-1)}\,.
\ee
This is a good approximation in our case since both the sonic and the Alfv\'enic Mach numbers (Eqs.~\ref{Mson} and \ref{Malf}, 
respectively) are
greater than 1.
The value of the parameter $\lambda$ depends on the shock efficiency $\eta$ and it reads
\be\label{shockeff}
\eta=\frac{4}{3\sqrt{\pi}}(r-1)\lambda^{3}e^{-\lambda^{2}}\,.
\ee
Equations~(\ref{nCR})--(\ref{I2}) are general and can be adapted to any acceleration scenario presented in
Sect.~\ref{Sect2}. 
  
\section{Potential CR acceleration sites}\label{fulfilment}
In this section, we identify and characterise possible sites of CR acceleration in protostars.
In particular, we consider accretion flows in the envelope and on the protostellar surface of Class~0 objects 
and also jets in more evolved sources.

\subsection{Accretion flows in collapsing envelopes}\label{accflows}
A number of Class~0 collapsing envelopes have been observed and their density and temperature profiles have been modelled
(e.g. Ceccarelli et al.~\cite{cc00}; Crimier et al.~\cite{cc09}).
Assuming a spherical collapse, $U\simeq1-10$~km~s$^{-1}$ at 100~AU.
If $B=10~\mu$G in the initial (pre-collapse) volume of radius 0.1~pc and assuming field freezing,
the magnetic field strength in the final
(post-collapse) volume of radius 100~AU is about 400~mG.
This naive estimate is comparable within a factor of 2 with the value found by Alves et al.~(\cite{av12}) 
who estimate $B\approx200$~mG from 
observations of shock-induced H$_{2}$O masers. This is an averaged quantity, but Imai et
al.~(\cite{in07}) computed the position of the H$_{2}$O masers, and found the farther one to be
at about 110~AU. Finally, the ionisation fraction has to be of the order of $10^{-4}$ to $10^{-5}$
in order to justify the presence of maser pumping (Strelnitskij~\cite{s84}; Wootten~\cite{w89}).
Masers arise in the presence of shocks and  they are usually associated with jet activity rather than accretion flows; in other words, the values for both magnetic field strength and ionisation fraction have to be regarded as upper limits in our estimates.
We check all the conditions in Sect.~\ref{conditionsDSA} and make a parameter study using the ranges of 
values shown in the first row
of Table~\ref{paramspace} verifying that Eq.~(\ref{Rratio}) 
is not fulfilled in accretion flows (${\mathscr R}\ll1$). The ionisation fraction and the
shock velocity are too small and they quench the CR acceleration, and
the magnetic field strength is also  high enough to produce a sub-Alfv\'enic shock.
This means that we can rule out accretion flows as possible CR acceleration
sites.
\begin{table}[!ht]
\setlength{\tabcolsep}{3pt}
\caption{Ranges of values of the parameters described in the text: upstream flow velocity in the shock reference frame ($U$),
temperature ($T$), total hydrogen density ($n_{\rm H}$), ionisation fraction ($x$), and magnetic field strength ($B$).}
\begin{center}
\begin{tabular}{cccccc}
\hline\hline
site$^{*}$ & $U$ & $T$ & $n_{\rm H}$ & $x$ & $B$\\
     & $\mathrm{[km~s^{-1}]}$ & $\mathrm{[K]}$ & $\mathrm{[cm^{-3}]}$ & & $\mathrm{[G]}$\\ 
\hline
${\mathscr E}$ & $1-10$ & $50-100$ & $10^{7}-10^{8}$ & $\lesssim10^{-6}$ & $10^{-3}-10^{-1}$\\
${\mathscr J}$ & $40-160$ & $10^{4}-10^{6}$ & $10^{3}-10^{7}$ & $0.01-0.9$ & $5\times10^{-5}-10^{-3}$\\
${\mathscr P}$ & 260 & $9.4\times10^{5}$ & $1.9\times10^{12}$ & $0.01-0.9$ & $1-10^{3}$\\
\hline
\end{tabular}\\
$^{*}{\mathscr E}$ = envelope (Sect.~\ref{accflows}); 
${\mathscr J}$ = jet (Sect.~\ref{jet});\\ 
${\mathscr P}$ = protostellar surface (Sect.~\ref{stellarsurfaceacc}).
\end{center}
\label{paramspace}
\end{table}
\normalsize

\subsection{Jets}\label{jet}
Jets are observed at all stages during the evolution of a protostar, from the main infall phase of Class~0 objects
(e.g. HH~212, McCaughrean et al.~\cite{mz02}) to evolved Class~I protostars (e.g. HH~111, Reipurth et al.~\cite{rh97}) and
to Class~II sources (e.g. HH~30, Watson \& Stapelfeldt~\cite{ws04}).
Jet speeds, $v_{\rm jet}$, are similar for different classes, between about 60 and 300~km~s$^{-1}$ with shock velocities, $v_{\rm sh}$, 
of the order
of $20-140$~km~s$^{-1}$ (Raga et al.~\cite{rv02,rn11}; Hartigan \& Morse~\cite{hm07}; Agra-Amboage et al.~\cite{ad11}).
In the equations of Sect.~\ref{conditionsDSA}, $U$ is the upstream flow velocity in the shock reference frame
(see Eq.~\ref{velref} with $v_{\rm fl}=v_{\rm jet}$). Taking the extreme values of  $v_{\rm jet}$ and $v_{\rm sh}$, we assume  $U$ to be in the
range $40-160$~km~s$^{-1}$.
A stationary shock is seen at 20~AU in Class~I and II protostars, 
while for the time being the  resolution is too low for  Class~0 objects.
There are also moving internal shocks, spaced each other by about 100 AU.

The total hydrogen density is between $10^{3}$ and $10^{7}$~cm$^{-3}$ (Lefloch et al.~\cite{lc12}; G\'omez-Ruiz et al.~\cite{gg12})
with temperatures from about $10^{4}$~K up to about $10^{6}$~K (Frank et al.~\cite{fr14}).
Thus far there are no measurements of magnetic field strengths. The only theoretical estimate has been carried out by
Te\c sileanu et al.~(\cite{tm09,tm12}) who have found $B\sim300-500~\mu$G for Class~II sources.
The transverse radius of a jet is about 10~AU
and 50~AU at a distance of 100~AU and 1000~AU from the source, respectively,
and the opening angle spans from about $4^{\circ}$ for RW~Aur to about $15^{\circ}$ for DG~Tau (Cabrit et al.~\cite{cc07}). 
Hartigan et al.~(\cite{he04}) give
estimates closer to the source: about 5~AU of transverse radius at a distance of about 15~AU from the central object for the two 
Classical T Tauri stars HN~Tau and UZ~Tau~E.

The ionisation fraction in Class~II objects for the atomic gas can be as high as 0.9,\footnote{
If $T\approx10^{6}$~K, then $x\approx1$.}
decreasing towards the source as a result  of recombination
processes due to higher densities (Maurri et al.~\cite{mb14}). The ionisation fraction in Class~I objects
is similar to that in Class~II objects,
$x\sim0.05-0.9$, but with higher electron and total densities (Nisini et al.~\cite{nb05}; 
Antoniucci et al.~\cite{an08}; Garcia L\'opez et al.~\cite{gn08}; Frank et al.~\cite{fr14}).
Conversely, Class~0 jets are mainly molecular, allowing for 
a rapid dissociative recombination that acts  to dramatically decrease
the ionisation fraction. 
For instance in the Class~0 HH~211, $n_{\rm H}\sim10^{5}$~cm$^{-3}$ with $x\sim(1.6-5)\times10^{-3}$ at about
1000~AU from the source (Dionatos et al.~\cite{dn10}).
The second row of Table~\ref{paramspace} summarises the range of parameter values.

\subsection{Protostellar surfaces}
\label{stellarsurfaceacc}
In order to study the efficiency of CR acceleration on protostellar surfaces, 
we used the computational results of Masunaga \& Inutsuka~(\cite{mi00}) for the Class~0 protostellar collapse of an initially homogeneous 
cloud core. Their simulation describes the phase of main accretion
when the protostar mass grows because of the steady accretion from the infalling envelope.
They give the temporal evolution of temperature, density, and flow velocity in the observer reference frame, 
which --
assuming a stationary shock -- is equal
to the upstream flow velocity in the shock reference frame. 
Considering the shock to be stationary, the downstream region is close to the protostar and the upstream
region towards the envelope, namely the opposite configuration with respect to the jet shock case.
The radius of the protostar is set to $2\times10^{-2}$~AU and we find that only the last time step of the simulation, 
corresponding to the 
end of the main accretion phase, leads to a strong proton acceleration. 
The ranges of parameter values are listed in the third row 
of Table~\ref{paramspace}.

\section{Spectrum of accelerated CRs at the shock surface}\label{spectra}

For jet and protostellar surface shocks 
we perform a parametric study using the values in the second and third rows of Table~\ref{paramspace}.
In order to calculate $\widetilde{P}_{\rm CR}$, we fix $\eta=10^{-5}$ in Eq.~(\ref{eqPcr}) and we omit
the contribution of the last term, which contains $\widetilde{p}_{\rm max}$.
To be consistent,  $E_{\rm max}$  should be recursively computed and  then  Eq.~(\ref{eqPcr}) solved, but we verified that the variation
in $\widetilde{P}_{\rm CR}$ is lower than a factor of about 3. With this assumption, $\widetilde{P}_{\rm CR}$ only depends on
the shock velocity with respect to the upstream flow and the ionisation fraction,
spanning from $2\times10^{-3}$ ($U=40~\mathrm{km~s^{-1}}$, $x=0.9$) to $5\times10^{-2}$ 
($U=160~\mathrm{km~s^{-1}}$, $x=0.01$).

We study the case of a Bohm-type diffusion shock ($k_{\rm u}=1$, see Eq.~\ref{diffusioncoeff}) 
since this is the most favourable circumstance for accelerating 
CRs in the case of self-generated waves (see also Sect.~\ref{keffects}). 
In fact, the upstream diffusion coefficient can be written as a function of the magnetic field strength
and of its turbulent component, $\delta B$
(see Drury~\cite{d83}):
\be\label{kukbdeltabb2}
k_{\rm u}=\left(\frac{\kappa_{\rm u}}{\kappa_{\rm B}}\right)^{-\alpha}=\left(\frac{B}{\delta B}\right)^2\,.
\ee
If $k_{\rm u}=1$, then $\delta B=B$, i.e. the magnetic fluctuations responsible for pitch angle scattering are  large enough to cause DSA to be effective at its maximum degree (see also Sect.~\ref{conditionsDSA}).
To justify this assumption, we compute $k_{\rm u}$ following Pelletier et al.~(\cite{pl06}) for the case of a parallel shock ($\alpha=-1$)
\begin{eqnarray}
k_{\rm u}&=&\frac{2}{\widetilde{P}_{\rm CR}}\frac{V_{\rm A}}{U}=%
4\times10^{-2}\left(\frac{U}{10^{2}~\mathrm{km~s^{-1}}}\right)^{-1}\\\nonumber%
&&\times
\left(\frac{n_{\rm H}}{10^{6}~\mathrm{cm^{-3}}}\right)^{-0.5}\left(\frac{B}{10~\mathrm{\mu G}}\right)%
\left(\frac{\widetilde{P}_{\rm CR}}{10^{-2}}\right)^{-1}\,.
\end{eqnarray}
With the values in the second and third rows of Table~\ref{paramspace},
$k_{\rm u}$ is found to be about 1.

\subsection{Maximum CR energy at the jet shock surface}\label{accCRsatjetsurface}

For jet shocks, we set the temperature at $T=10^{4}$~K, and consider 
magnetic field strengths between 50~$\mu$G and 1~mG as well as shock velocities between
40 and 160~km~s$^{-1}$, then we study the parameter space of
hydrogen density, \mbox{$n_{\rm H}\in[10^{3},10^{7}]$~cm$^{-3}$}, and ionisation fraction, $x\in[0.01,0.9]$, for a shock at a distance $R_{\rm sh}=100$~AU from the protostar and a transverse
radius $R_{\perp}=10$~AU. 

\begin{figure}[!t]
\begin{center}
\resizebox{\hsize}{!}{\includegraphics{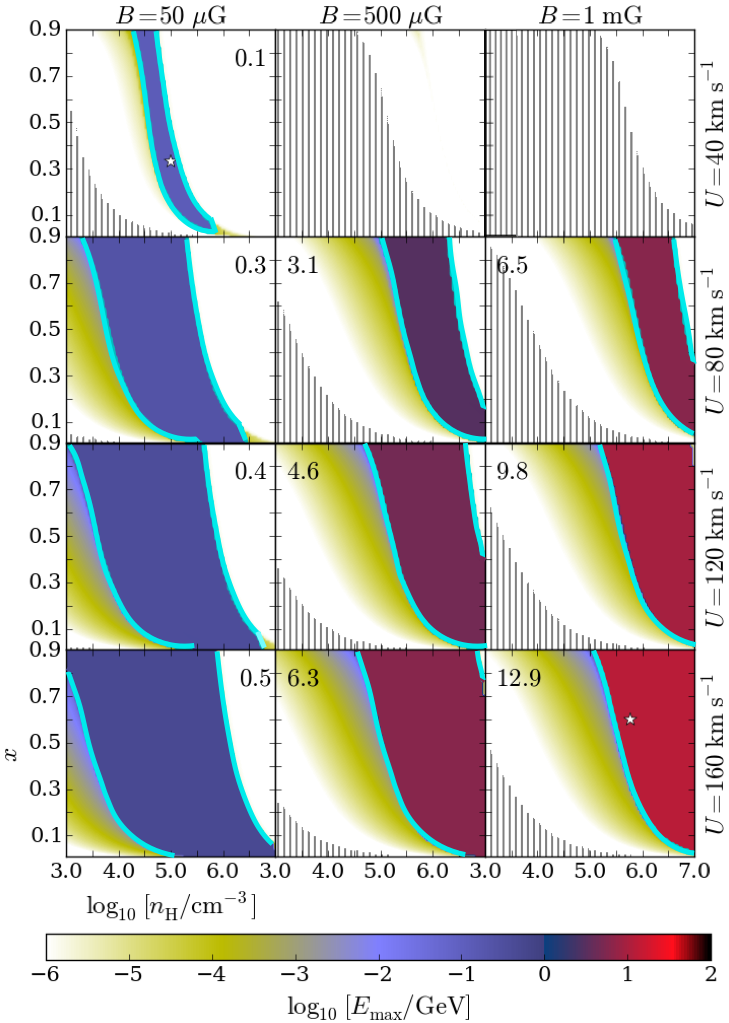}}
\caption{Case of a parallel shock in jets: 
ionisation fraction, $x$, versus total hydrogen density, $n_{\rm H}$, 
for different combinations of initial parameters. The magnetic field strength varies from
50~$\mu$G to 1~mG (from left to right), while the shock velocity varies from 40 to 160~$\mathrm{km~s^{-1}}$ (from top to bottom). 
The temperature is assumed
equal to $10^{4}$~K and the shock distance from the protostar and its transverse radius are $R_{\rm sh}=100$~AU and 
$R_{\perp}=10$~AU, respectively. The colour map shows the values of $E_{\rm max}$ in the case of a parallel shock
when the conditions imposed by
Eqs.~(\ref{Uhighercsva}) and~(\ref{Rratio}) are simultaneously satisfied.
{\em Cyan} contours delimit the regions where $E_{\rm max}$ reaches its maximum asymptotic value
shown in GeV in each subplot. Vertically hatched regions refer to combinations of parameters corresponding 
to strong wave damping ($\mathscr{R}<1$).
The two {\em solid white stars} in the upper left and lower right plots show the values of $n_{\rm H}$ and $x$ considered for
the evaluation of the emerging spectra (models \mW\ and \mS\ in Fig.~\ref{emergingspectra}).}
\label{nxT4}
\end{center}
\end{figure}

We first consider the case of a parallel shock.
Figure~\ref{nxT4}
shows the maximum energy that a shock-accelerated proton can reach in the case of a parallel shock.
We find that for low shock velocities with respect to the upstream flow
($U=40~\mathrm{km~s^{-1}}$) the acceleration process
is efficient only for low values of $B$ (for $B>100~\mu$G, $E_{\rm max}$ is lower than $1-10$~keV). 
By increasing both $U$ and $B$, the maximum energy
attains higher values (from about 100~MeV to about 13~GeV). 
It is interesting to note that once the combination of parameters satisfies the condition $\mathscr{R}>1$ (see Eq.~\ref{Rratio}),
$E_{\rm max}$ rapidly reaches a constant value, encompassed by the cyan contours in each subplot. 
In fact, the maximum energy is generally controlled by
downstream escape (Eq.~\ref{FEescd}), which is independent of both $n_{\rm H}$ and $x$.
The number in each subplot shows the maximum value of $E_{\rm max}$ in GeV that can be attained for given
values of the shock velocity and the magnetic field strength. 

Electrons can be accelerated as well, but generally $E_{\rm max}$ for electrons is much lower than $E_{\rm max}$ for protons because
of wave damping and stronger energy losses. For instance, for $U=160$~km~s$^{-1}$ and $B=1~\mathrm{mG}$, $E_{\rm max}\sim300$~MeV for 
a narrow range of density and ionisation fractions ($n_{\rm H}\gtrsim3\times10^{6}$~cm$^{-3}$, $x\gtrsim0.6$). For lower values of 
$B$ and $U$, $E_{\rm max}\lesssim50$~MeV.

Supposing the magnetic field to have a strong toroidal component, we repeat the calculation for the case of a perpendicular shock. We find
that $E_{\rm max}$ increases by a factor of \mbox{$E_{\rm esc,d}^{\perp}/E_{\rm esc,d}^{\parallel}= r(r+1)/2$}, where
superscripts $\perp$ and $\parallel$ refer to a perpendicular and parallel shock, respectively.

For the sake of completeness, we show in Sect.~\ref{caseT5}, Fig.~\ref{nxT5}, the results for a higher temperature, $T=10^{5}$~K. 

\subsection{Accelerated CR spectrum at the protostellar shock}

\begin{figure}[!ht]
\begin{center}
\resizebox{\hsize}{!}{\includegraphics{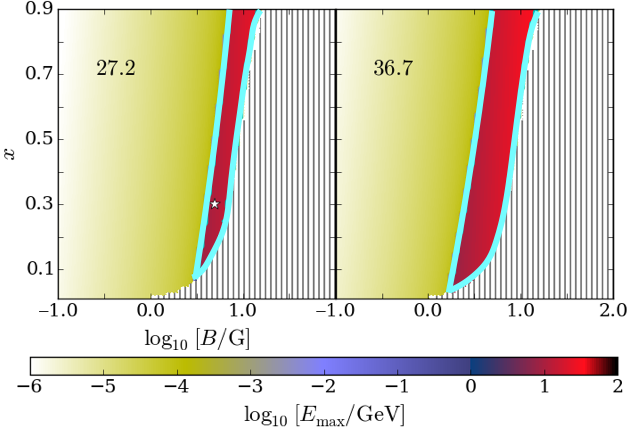}}
\caption{
Case of parallel ({\em left}) and perpendicular ({\em right}) shocks on protostellar surfaces:
ionisation fraction, $x$, versus magnetic field strength, $B$, 
for the parameters specified in the third line of Table~\ref{paramspace} ($R_{\rm sh}=2\times10^{-2}$~AU). 
The colour map shows the values of $E_{\rm max}$ when the conditions imposed by
Eqs.~(\ref{Uhighercsva}) and~(\ref{Rratio}) are simultaneously verified. 
The {\em  cyan} contour delimits the region where $E_{\rm max}$ gets its maximum asymptotic value
shown in GeV in each subplot. Vertically hatched regions refer to combinations of parameters corresponding 
to strong wave damping ($\mathscr{R}<1$).
The {\em solid white star} in the left panel shows the values of $B$ and $x$ considered for
the evaluation of the emerging spectrum (models \mP\ in Fig.~\ref{emergingspectra}).}
\label{stellarsurfaceplot}
\end{center}
\end{figure}

For protostellar surfaces, we study the parameter space of magnetic field strength and ionisation fraction using $\epsilon=0.1$ (Eq.~\ref{FEescu}), 
$\eta=10^{-5}$ (Eq.~\ref{eqPcr}), and
$k_{\rm u}=1$ (Eq.~\ref{kukbdeltabb2}) for both a parallel and a perpendicular shock. 
Figure~\ref{stellarsurfaceplot} shows that at the protostellar surface
accelerated CR protons can reach $E_{\rm max}\approx26$~GeV and $E_{\rm max}\approx37$~GeV in the case of a parallel and
a perpendicular shock, respectively. 
The values of the magnetic field where CR acceleration
is possible, $B\sim3-10$~G, are compatible with those computed by e.g. Garcia et al.~(\cite{gf01}).
Because of high temperatures, Coulomb losses are dominant and $E_{\rm max}$ is constrained by $E_{\rm loss}$. Thus, for a
perpendicular shock $E_{\rm max}$ increases by a factor of $E_{\rm loss}^{\perp}/E_{\rm loss}^{\parallel}=
(r+1)/2$. 

The maximum energy can be even higher in the case of massive protostars where shocks are much stronger because the mass is higher.
It is important to bear in mind that, in principle, CRs accelerated at the protostellar surface shock could also be re-accelerated in jet shocks.

\subsection{Emerging CR spectrum at the shock surface}
The solution given by Eq.~(\ref{fp}) is used to compute the emerging CR spectrum at the shock surface. 
The energy distribution function of shock-accelerated CRs reads
\be\label{Nsh}
{\mathscr N}(E)=4\pi p^{2}f(p) \frac{\ud p}{\ud E}\,.
\ee
Then, the CR flux emerging from the shock surface, $j(E)$, 
namely the number of particles per unit energy, time, area, and solid angle
reads
\be\label{jCR}
j(E)=\frac{v(E)\mathscr{N}(E)}{4\pi}\,.
\ee

We compute the emerging CR proton spectrum in the case of a parallel shock both in a jet (for two opposite and extreme cases:
a weak and a strong shock, labelled \mW\ and \mS, respectively) and on a protostellar surface, labelled \mP.
The relevant parameters
are shown in Table~\ref{paramf0} and by the white stars in Figs.~\ref{nxT4} and \ref{stellarsurfaceplot}.
We note that when assuming $\eta=10^{-5}$, the normalised CR pressure (Eq.~\ref{Pcrprime}) for all three models is lower than 10\%, as predicted in Sect.~\ref{PCR}. 
\begin{table*}[!th]
\caption{Parameters to calculate the particle distribution $f(p)$ in the case of parallel shocks 
for $\kappa_{\rm u}=\kappa_{\rm B}$ and $\eta=10^{-5}$.}
\begin{center}
\begin{tabular}{cccccccccccc}
\hline
\vspace{-.3cm}
\\
model & $U$ & $B$ & $n_{\rm H}$ & $x$ & T & $r$ & $E_{\rm max}$ & $\widetilde{P}_{\rm CR}$ & $\lambda$ & $p_{\rm inj}$ & $p_{\rm max}$\\
&$[\mathrm{km\ s^{-1}}]$ & $[\mathrm{G}]$ & $[\mathrm{cm^{-3}}]$ & & [$10^{4}$~K] &  & [GeV]  & [$10^{-2}$] & & [MeV/c] & [GeV/c]\\
\hline\hline
${\mathcal W}$ & 40 &$5\times10^{-5}$& $10^{5}$ & 0.33  & 1 & 2.977 & 0.13 & 0.88 & 4.010 & 0.306 & 0.505\\
${\mathcal S}$ & 160&$10^{-3}$       & $6\times10^{5}$ & 0.60 & 1 & 3.890 & 12.9 & 4.70 & 4.062 & 1.146 & 13.762\\
${\mathcal P}$ & 260& 5              & $1.9\times10^{12}$ & 0.30 & 94 & 2.290 & 11.4 & 0.03 & 3.950 & 2.058 & 12.306\\ 
\hline
\end{tabular}
\end{center}
\label{paramf0}
\end{table*}

\begin{figure}[!ht]
\begin{center}
\resizebox{\hsize}{!}{\includegraphics{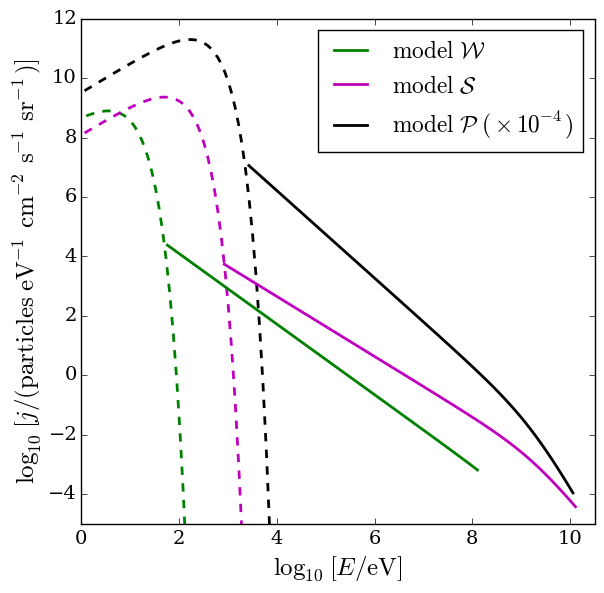}}
\caption{Emerging spectra of the shock-accelerated protons ({\em solid lines}) for the models described in the text.
The {\em dashed lines} represent the corresponding Maxwellian distributions of the thermal protons.}
\label{emergingspectra}
\end{center}
\end{figure}

Figure~\ref{emergingspectra} displays the above shock-accelerated proton spectra 
together with their corresponding Maxwellian distributions of thermal protons.

\subsection{Effect of specific parameters on $E_{\rm max}$}
There are a number of mechanisms that can dramatically attenuate the emerging CR spectrum at the shock surface
such as variations in diffusion coefficient (Sect.~\ref{keffects}), in CR pressure (Sect.~\ref{CRpressureeffects}),
and in temperature (Sect.~\ref{caseT5}).

\subsubsection{Upstream diffusion coefficient}\label{keffects}
\begin{figure}[!t]
\begin{center}
\resizebox{\hsize}{!}{\includegraphics{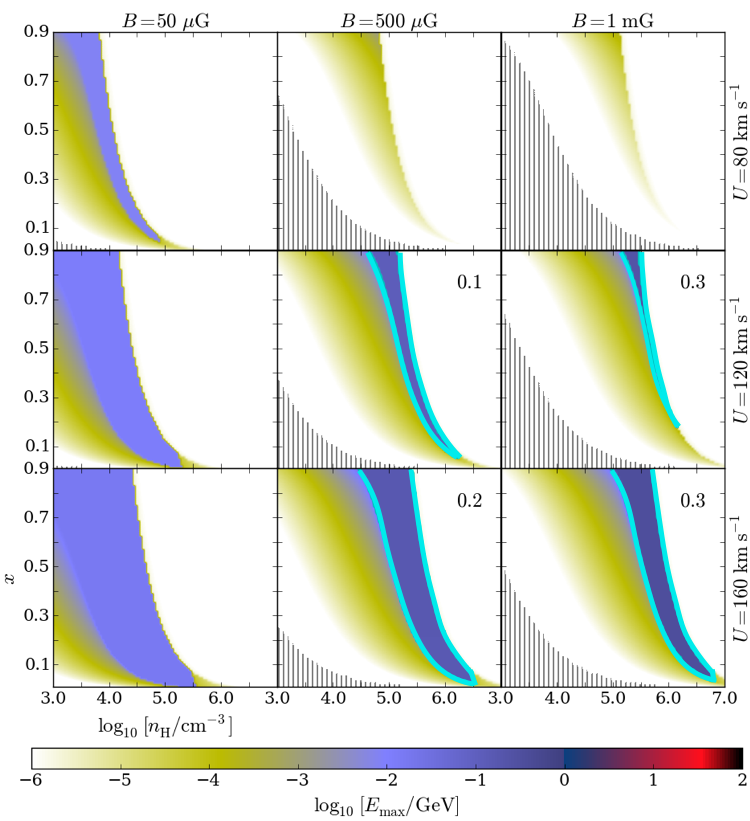}}
\caption{Same as Fig.~\ref{nxT4}, but for an upstream diffusion coefficient \mbox{$\kappa_{\rm u}=30\,\kappa_{\rm B}$}
for a parallel shock.}
\label{nxT4ku30}
\end{center}
\end{figure}

\begin{figure}[!h]
\begin{center}
\resizebox{\hsize}{!}{\includegraphics{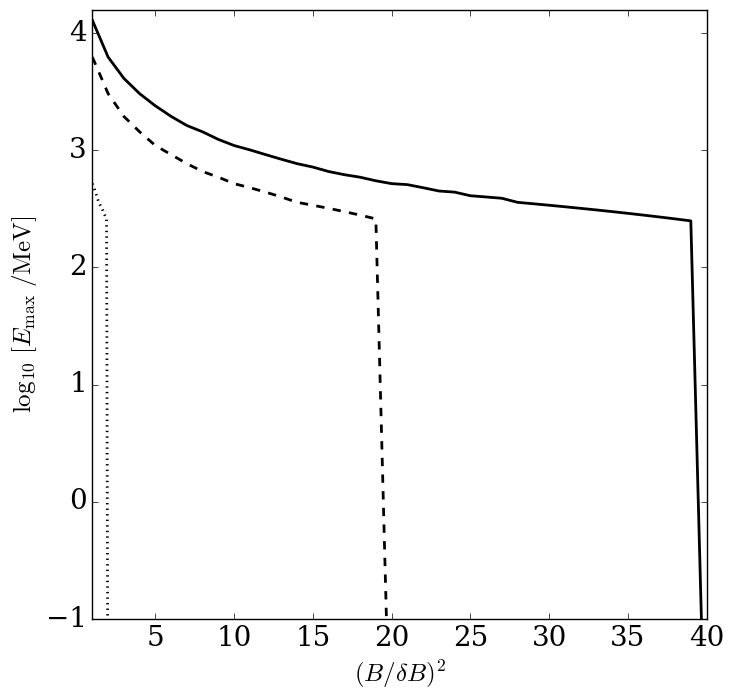}}
\caption{Maximum energy of accelerated protons as a function of $(B/\delta B)^{2}$ in the case of parallel shocks 
for 
$U=160~\mathrm{km~s^{-1}}$, $n_{\rm H}=6\times10^{5}~\mathrm{cm^{-3}}$, and $x=0.6$ with $B=50~\mu$G ({\em dotted line}), 
500~$\mu$G ({\em dashed line}), and 1~mG ({\em solid line}).}
\label{emax_vs_ku}
\end{center}
\end{figure}

A variation of the diffusion coefficient can strongly modify the CR flux.
In our derivation, we assume Bohm-like diffusion ($\kappa_{\rm u}=\kappa_{\rm B}$), but --
since $\kappa_{\rm u}=\kappa_{\rm B}(B/\delta B)^{-2\alpha}$ (Eq.~\ref{kukbdeltabb2}) --
for a parallel shock $\kappa_{\rm u}\geq\kappa_{\rm B}$, while for a perpendicular shock \mbox{$\kappa_{\rm u}\leq\kappa_{\rm B}$}.
In the case of parallel shocks, an increase in $\kappa_{\rm u}$ corresponds to a reduction in  $\delta B$, 
the turbulence produced by the accelerated CRs that is responsible for DSA
(see Sect.~\ref{conditionsDSA}), resulting in a decrease in the shock acceleration efficiency.
As shown in Fig.~\ref{nxT4ku30}, 
considering for instance $\kappa_{\rm u}=30\,\kappa_{\rm B}$ for a parallel shock, the shock velocity has to be
higher than at least 80~km~s$^{-1}$ and the magnetic field greater than 50~$\mu$G 
to accelerate CR protons above 100~MeV.
In order to assess how much the upstream diffusion coefficient 
affects $E_{\rm max}$, we compute the relation between these two quantities
for a parallel shock,
assuming $U=160~\mathrm{km~s^{-1}}$, $n_{\rm H}=6\times10^{5}$~cm$^{-3}$, and $x=0.6$,
such as in model ${\cal S}$.
As shown in Fig.~\ref{emax_vs_ku}, the values of $k_{\rm u}=(B/\delta B)^{2}$ at which 
$E_{\rm max}$ drops below the threshold for efficient acceleration are
about 2, 20, and 40 for $B=50~\mu$G, 500~$\mu$G, and 1~mG, respectively.

In the case of perpendicular shocks, if $\kappa_{\rm u}$ decreases, then $\delta B$ also decreases. However, as pointed out by 
Jokipii~(\cite{j87}), in this case $E_{\rm max}$ increases by a factor of $k_{\rm u}k_{\rm d}=(B/\delta B)^{4}/r$ 
(see also Eq.~\ref{FEescd}). In  this configuration particles drift along the shock face colliding with it several times in a single
scattering mean free path. In other words, since the magnetic field turbulence decreases as does  the particle-wave scattering,
particles are more easily caught by the shock, the acceleration time is reduced, and then $E_{\rm max}$ increases.
Nevertheless, two effects limit the increase in $E_{\rm max}$. First, the perpendicular transport is usually controlled by magnetic field
line wandering (Kirk et al.~\cite{kd96}), 
and it is enhanced with respect to the solution obtained from pure scattering, and the expected $E_{\rm max}$ is reduced.
Second,  
to avoid any anisotropy in the particle distribution, in order for DSA to take place at the injection momentum, $p_{\rm inj}$
(Eq.~\ref{injmom}),
particles must be scattered in the time
required to drift through the shock (Jokipii~\cite{j87}). 
This gives a constraint to the maximum value of $k_{\rm u}$, $k_{\rm u,max}$, which in turn
defines $E_{\rm max}$,
\be
k_{\rm u,max}=\frac{\beta_{\rm inj} c}{U}\,,
\ee
where $\beta_{\rm inj}$ is related to $p_{\rm inj}$.
Once the temperature is fixed, the injection momentum only depends on the shock velocity with respect to the upstream flow, then 
$k_{\rm u}$ is uniquely a function of $U$. For instance, using the range of $U$ for jets (see Sect.~\ref{jet}), 
$k_{\rm u,max}$ is  limited between 2.44 ($U=40$~km~s$^{-1}$) and 2.29 ($U=160$~km~s$^{-1}$).
This means that for a perpendicular shock, DSA is efficient in a narrow range of $k_{\rm u}$, 
but particles can  still be injected in the acceleration process
by means of other mechanisms (see Appendix~\ref{app:othermechanisms}).
However, for $U=160$~km~s$^{-1}$, assuming $k_{\rm u}=k_{\rm u,max}=2.29$, $E_{\max}$ is about 100~GeV for
a shock at $R_{\rm sh}=100$~AU from the propostar. This means that for perpendicular shocks at larger $R_{\rm sh}$, where 
the transverse radius, $R_{\perp}$, is also larger, CRs can reach TeV energies and their $\gamma$ emission 
could be a target for Cherenkov telescopes (a work is in preparation to quantify this aspect).

\subsubsection{CR pressure}\label{CRpressureeffects}
The accelerated CR spectrum at the shock surface can also be attenuated if the CR pressure decreases. 
In fact, the normalisation constant $f_{0}$ (Eq.~\ref{partpress}) is
directly proportional to $P_{\rm CR}$, which in turn depends on the parameter $\eta$ (Eq.~\ref{eqPcr}).
In addition, CR pressure controls $E_{\rm max}$ through $E_{\rm damp}$ (Eq.~\ref{Edamp}). For instance, if in model \mS\ we decrease
$\eta$ by a factor of 10, then $E_{\rm max}$ decreases by a factor of about 500. This translates into a lower number of high-energy
CRs available to ``refill'' the low-energy part of the spectrum during propagation, so that thermalisation occurs
at a lower column density and the ionisation rate decreases.

\subsubsection{Temperature}\label{caseT5}
\begin{figure}[!t]
\begin{center}
\resizebox{\hsize}{!}{\includegraphics{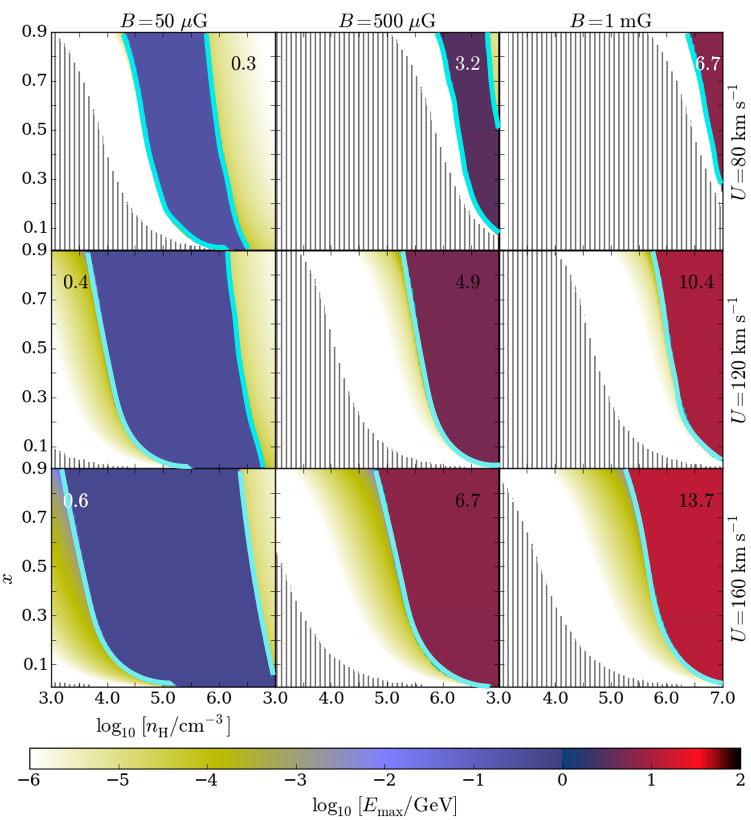}}
\caption{Same as Fig.~\ref{nxT4}, but for a temperature of 10$^{5}$~K.}
\label{nxT5}
\end{center}
\end{figure}

\begin{figure}[!t]
\begin{center}
\resizebox{0.8\hsize}{!}{\includegraphics{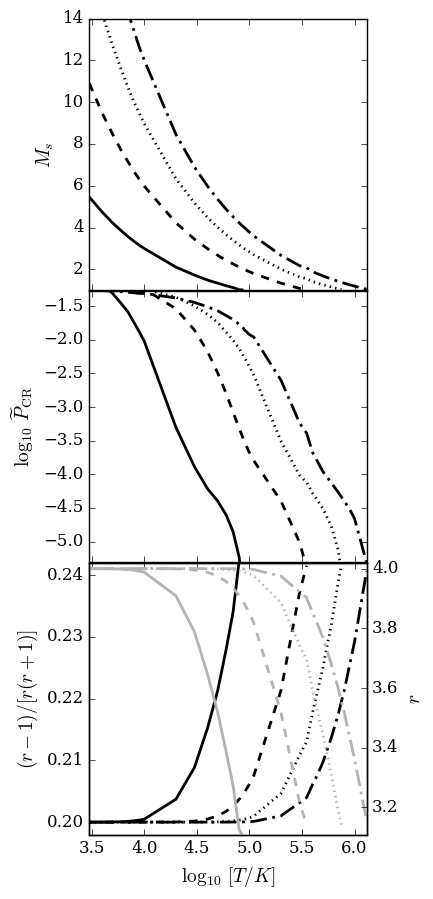}}
\caption{Sonic Mach number (upper panel), normalised CR pressure (middle panel),
compression ratio
and factor proportional both to $E_{\rm esc,d}$ and $E_{\rm loss}$ for parallel shocks 
({\em grey} and {\em black lines}, respectively, in lower panel) as a function of the temperature for 
$x=0.3$, $\eta=10^{-5}$;
\mbox{$U=40~\mathrm{km~s^{-1}}$} ({\em solid lines}),
\mbox{$U=80~\mathrm{km~s^{-1}}$} ({\em dashed lines}),
\mbox{$U=120~\mathrm{km~s^{-1}}$} ({\em dotted lines}), and
\mbox{$U=160~\mathrm{km~s^{-1}}$} ({\em dash-dotted lines}).}
\label{MsPcr_vs_T}
\end{center}
\end{figure}

An increase in temperature of one order of magnitude from $10^{4}$~K to $10^{5}$~K 
results in an almost negligible increase in the maximum 
energy achieved (see Fig.~\ref{nxT5}) because $E_{\rm max}$ is set by $E_{\rm esc,d}$ (Eq.~\ref{FEescd}) 
or $E_{\rm loss}$ (Eq.~\ref{FEloss}) for a jet shock or a protostellar surface shock,
respectively. Both $E_{\rm esc,d}$ and $E_{\rm loss}$, for parallel shocks, are proportional to $(r-1)/[r(r+1)]$ and the compression ratio $r$
depends
on the temperature (see lower panel of Fig.~\ref{MsPcr_vs_T}). 
In addition, the space of solutions narrows and there are no
combinations of total hydrogen density and ionisation fraction allowing the particle acceleration for $U=40$~km~s$^{-1}$.
In fact, the condition ${\mathscr R}>1$ (Eq.~\ref{Rratio}) is never satisfied because of the temperature dependence in the factor $\Xi$ 
(Eq.~\ref{XI}). The higher the temperature, the lower both the sonic Mach number and the particle pressure (see Fig.~\ref{MsPcr_vs_T}).
Then, for increasing temperatures the shock enters the subsonic regime, the condition in Eq.~(\ref{Uhighercsva})
is no longer fulfilled, and the acceleration process becomes ineffective. However, even when $M_{s}>1$, $\widetilde{P}_{\rm CR}$ can be so low that particle acceleration is damped ($\mathscr{R}<1$).

\section{Propagation of accelerated CRs in the jet}\label{energylossregime}
As shown in Fig.~\ref{sketch}, CRs are accelerated downstream of the shock surface ({\em acceleration zone}). Here the turbulence 
produced by CRs ($\delta B\lesssim B$) triggers the acceleration process. 
So far, $B$ and $\delta B$ have not been determined by observations, except for two sources where we have some information on 
the magnetic field morphology and magnitude (Carrasco-Gonz\'alez et al.~\cite{cr10}; Lee et al.~\cite{lr14}).

Moving farther and farther away from the shock, $\delta B$ decreases ({\em turbulence damping zone}) unless there are other turbulence sources.
In a partially ionised medium such as in a jet, damping
occurs through ion-neutral collisions for waves generated by particles with energies above a few MeV and through resonant interaction 
with the background plasma, i.e. linear Landau damping, for waves generated by particles with energies below a few MeV (see 
Appendix~\ref{app:dampingrate}).
It is found that the resonant self-generated waves produced at the shock front are damped over length scales much shorter than
the distance between the inner shock in the jet 
and the termination shock. 

\begin{figure}[!h]
\begin{center}
\resizebox{\hsize}{!}{\includegraphics{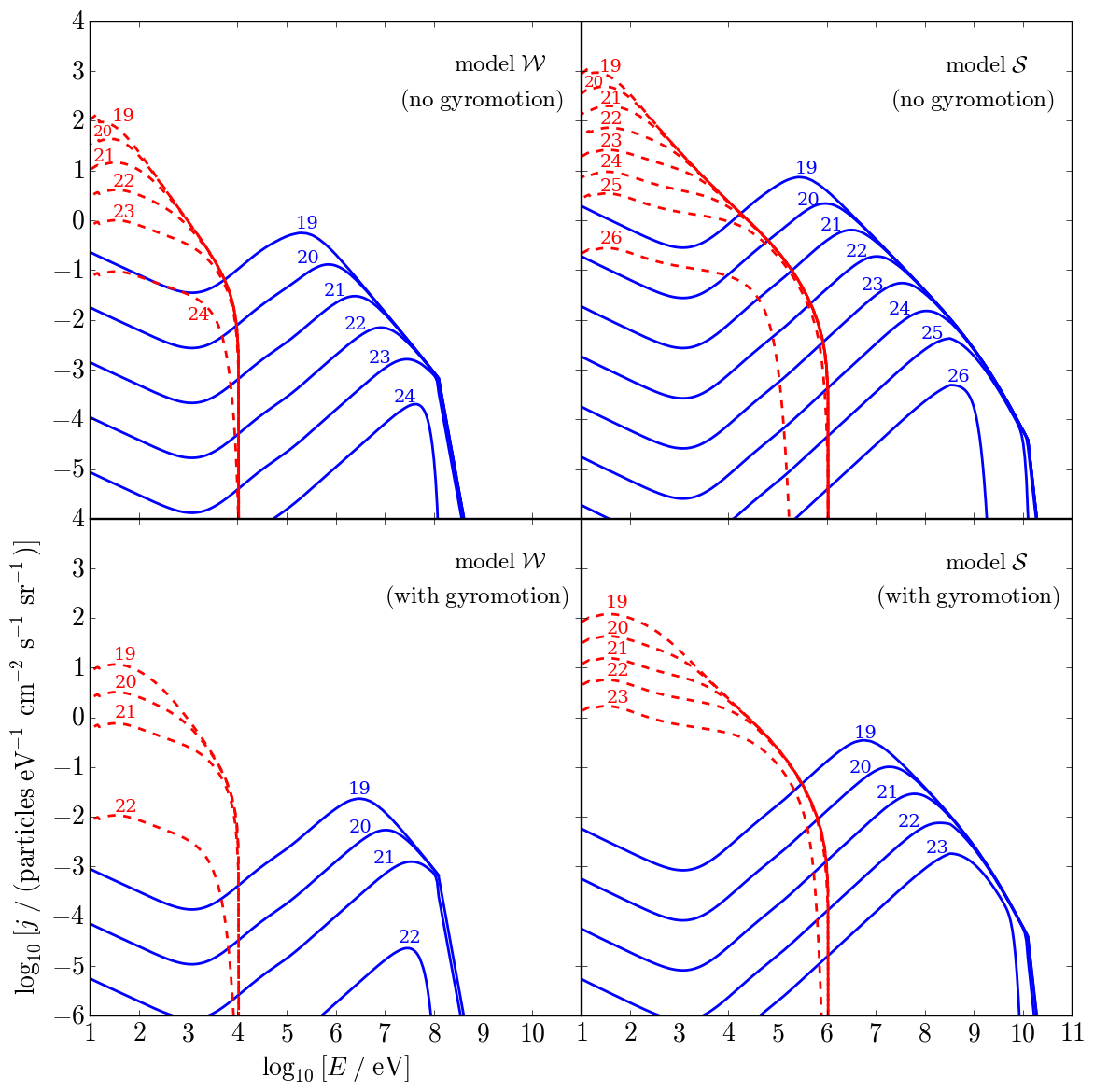}}
\caption{Propagated spectra in the absence of magnetic turbulence 
for models \mW\ ({\em left column}) and \mS\ ({\em right column})
neglecting and accounting for gyromotion effects ({\em upper} and {\em lower row}, respectively).
{\em Blue solid} and {\em red dashed lines} show the attenuated proton and secondary electron spectra, respectively, at increasing
depth in the cloud labelled by values of the column density along the line of sight, $\log_{10}\ [N_{\rm los}\mathrm{(H_{2})/cm^{-2}}]$.}
\label{modelWS_final}
\end{center}
\end{figure}

Entering the propagation zone, 
we assume that the turbulence created at the jet-outflow interface discussed in Appendix~\ref{shear} is negligible.
In this case, CR propagation in the jet is 
dominated by energy losses and can be treated using the continuous slowing-down approximation 
(Takayanagi~\cite{t73}; Padovani et al.~\cite{pgg09}).
Neglecting magnetic turbulence, we can imagine that CRs propagate by gyrating around magnetic field lines, losing energy
because of collisions with hydrogen atoms or molecules\footnote{Jets in Class~0 protostars are mainly molecular, while hydrogen is mostly in atomic form in more evolved sources.}.
We compute the attenuation of accelerated CR protons at the jet shock surface shown in Fig.~\ref{emergingspectra},
along the jet and towards the termination shock, using the method developed in Padovani et al.~(\cite{pgg09}).
The two upper panels of Fig.~\ref{modelWS_final} show the results for both models \mW\ and \mS\
(see Table~\ref{paramf0} for more details).
The accelerated CR protons soon start  to lose energy with increasing column density 
and their flux is attenuated; the most
energetic CRs are slowed down to 0.1--100~MeV, contributing to the bulk of the ionisation.
We note that even if CR electrons are not efficiently accelerated in the jet (see Sect.~\ref{ionneutralfriction}), they are created
as the product of hydrogen ionisation due to CR protons (See Appendix~B in Ivlev et al.~\cite{ip15}
for the calculation of the secondary electron spectrum).
It is evident that the CR spectrum related to 
model \mW\ is attenuated faster with
column density than the spectrum related to  model \mS. This happens because $E_{\rm max}$ for model \mW\ is about 100~MeV, while for
model \mS\ it is about 13~GeV (see Table~\ref{paramf0}), 
so that the latter model has a larger ``reservoir'' of high-energy CR protons, which gradually populate the 
low-energy part of the spectrum.
We also note that if $E_{\rm max}\lesssim10^{5}$ eV, 
the spectrum is completely thermalised as soon as the column density is of the order of $10^{19}$~cm$^{-2}$. 
This is to say that CR protons are not sufficiently accelerated to take part in the ionisation process
and this happens irrespective of the shape of the spectrum at low energies.

Nevertheless, these CR spectra have to be regarded as upper limits. 
In fact, Padovani \& Galli~(\cite{pg11}) and Padovani et al.~(\cite{phg13}) demonstrated how the calculation of the ionisation rate
cannot be
carried out without accounting for gyromotion effects due to the presence of magnetic fields. Since CRs perform helicoidal motion
around field lines, the effective column density that they pass through, $N_{\rm eff}$, is higher than the column density along the line of 
sight, $N_{\rm los}$.
Currently there is no observational estimate of the magnetic field strength and of its configuration in protostellar jets, but
it is possible to conjecture about the 
presence of a strong toroidal component such as that depicted by Te\c sileanu et al.~(\cite{tm14}). 
Furthermore, the angle between magnetic field lines and the disc surface has to be lower than $60^{\circ}$ in order to have a successful
jet launching (Blandford \& Payne~\cite{bp82}).
Using Eqs.~(19)--(24) from 
Padovani et al.~(\cite{phg13}) and assuming that the toroidal field component is larger than about 50\% of the total field, we estimate that 
$N_{\rm eff}$ can be a factor of about $100-300$ higher than $N_{\rm los}$. Because of this increase in column density,
the CR proton flux of both models is more rapidly thermalised at $\sim5\times10^{22}$~cm$^{-2}$ and $\sim8\times10^{23}$~cm$^{-2}$ 
for models \mW\ and \mS\, respectively (see lower
panels of Fig.~\ref{modelWS_final}). 

\subsection{(Re-)acceleration at the reverse bow shock}\label{reaccatrBS}
The jet morphology is far from being universally defined. 
Jet lengths 
spread over orders of magnitude
and usually there is not just a single final bow shock, but the innermost knots are all resolved into bow shocks due to a time-variable jet
emitting dense-gas bullets
(e.g. McCaughrean et al.~\cite{mz02}).
The situation is further complicated by jet angle variations due to precession 
(e.g. Devine et al.~\cite{db97}) or orbital motion (e.g. Noriega-Crespo et al.~\cite{nr11}).
For the sake of simplicity, we account for a single shock at 100~AU from the protostar (see Sect.~\ref{jet}), 
following CR propagation up to the reverse
bow shock (rBS).
Once the accelerated CRs reach the rBS, they are subjected to further acceleration before entering the hot spot region
(see Sect.~\ref{twozonemodel} and Fig.~\ref{sketch}).
In Sect.~\ref{conclusions} we will briefly discuss the dependence of $E_{\rm max}$ on $R_{\rm sh}$
and on multiple shocks.

We suppose that two processes take place at the rBS: the acceleration of thermal 
protons (electrons are not efficiently accelerated, see Sect.~\ref{accCRsatjetsurface})
and the re-acceleration
of CRs propagated from the jet shock.
Finally, we expect the bow shock to be much weaker than the rBS because of the interaction with the surrounding material
and we neglect any further CR acceleration.

Shocks developed along a jet can also re-accelerate a pre-existing population of accelerated CRs. 
Following Melrose \& Pope~(\cite{mp93}), if the momentum
distribution function of the CRs accelerated at the jet shock and propagated upstream of the rBS is $f_{\rm JS,prop}(p)$, 
the distribution function of the re-accelerated CRs, $f_{\rm JS,reacc}(p)$, reads
\be\label{freacc}
f_{\rm JS,reacc}(p) = q_{\rm rBS} \left(\frac{p}{R}\right)^{-q_{\rm rBS}} \int_{0}^{p/R} \xi^{\,q_{\rm rBS}-1} f_{\rm JS,prop}(\xi)\ud\xi\,,
\ee 
where $q_{\rm rBS}=3r_{\rm rBS}/(r_{\rm rBS}-1)$ is the shock index, $r_{\rm rBS}$ is the compression ratio at the rBS, and
$R^{3}=r_{\rm JS}$ accounts for adiabatic losses because of the decompression that develops behind the rBS.
It is important to note that re-acceleration at the rBS also involves the secondary CR electrons produced during primary ionisation.
These CR electrons are boosted up to relativistic energies and in Sect.~\ref{dgtau} we will show their relevance to explaining
synchrotron emission.

The total CR distribution function at the reverse bow shock surface, $f_{\rm rBS}(p)$, is given by
\be\label{ftot}
f_{\rm rBS}(p)=f_{\rm rBS,th}(p)+f_{\rm JS,reacc}(p)\,,
\ee
where $f_{\rm rBS,th}$ is the distribution function of thermal protons accelerated at the rBS surface and
$f_{\rm JS,reacc}$ is given by Eq.~(\ref{freacc}).
In order to compare the contribution of the re-accelerated CRs with the more freshly accelerated component of thermal protons at the rBS,
we consider for instance 
both models \mW\ and \mS\ and a distance for the rBS, $D_{\rm rBS}=1.8\times10^{3}$~AU, corresponding to
the position
of the bow shock in DG~Tau (knot C; Eisl\"offel \& Mundt~\cite{em98}).
Assuming a constant total hydrogen density of $10^{5}$~cm$^{-3}$ and $6\times10^{5}$~cm$^{-3}$ for model \mW\ and \mS, respectively
(see Table~\ref{paramf0}), the accelerated CRs pass through a
line-of-sight column 
density of $2.7\times10^{21}$~cm$^{-2}$ and $1.6\times10^{22}$~cm$^{-2}$, respectively.
The propagated CR proton and electron spectra, including gyromotion effects (see Sect.~\ref{energylossregime}), from the jet shock surface are labelled  ``JS,prop'' in Fig.~\ref{spectrumatrBS_paper_ok}.
The decrease at low energies is due to energy losses during the propagation (see also Fig.~\ref{modelWS_final}).
Using these spectra as input to the integral of Eq.~(\ref{freacc}), we compute the re-accelerated CR spectra at the rBS surface
(``JS,reacc'' in Fig.~\ref{spectrumatrBS_paper_ok}), making use of Eqs.~(\ref{Nsh}) and~(\ref{jCR}).

We evaluate the CR proton spectrum drawn from the thermal pool at the rBS 
(labelled ``rBS,th'' in Fig.~\ref{spectrumatrBS_paper_ok})
following Sect.~\ref{spectra} and assuming the same values for the
shock velocity, ionisation fraction, total hydrogen density, magnetic field strength, and temperature as for the shock at 100~AU. 
However, the considered transverse radius, $R_{\perp}$, which enters the evaluation of the maximum energy through the condition on 
shock geometry
(see Eqs.~\ref{FEescu} and~\ref{FEescd}), is larger by about two orders of magnitude 
than $R_{\perp}$ computed at 10~AU, and for DG~Tau it is about $1.3\times10^{3}$~AU.\footnote{$R_{\perp}$ at the bow shock of DG~Tau has been estimated from the value of the volume 
of the emitting region evaluated by Ainsworth et
al.~(\cite{as14}).}
As a consequence, $E_{\rm max}$ can increase  to TeV energies, being mainly constrained by downstream escape losses (see Sect.~\ref{jet}).

\subsection{Solution in the hot spot region}\label{twozonemodel}
After the rBS, in the hot spot region the flow is expected to be turbulent. The turbulence is likely connected with flow-ambient medium interactions. Cunningham et al.~(\cite{ck09}) (see references therein) considered the propagation of stellar jets in a turbulent medium that may be associated with a molecular cloud disrupted through thermal or Vishniac instability. 
Other fluid instabilities that can lead to turbulent motions are also expected while the jet propagates in the interstellar medium. 
Hence we account for the possibility that the downstream hot spot flow is turbulent, computing 
the CR distribution in the downstream medium using a two-zone model.
This approximation is strictly valid if the length scales over which the escape and loss processes occur are longer than the scale of the region under consideration. This allows us to use space-average diffusion and loss terms.

The particle energy distribution function
in the downstream jet medium, ${\mathscr N}(E,t)$, evolves following
\begin{equation}\label{Nevolve}
\frac{\partial}{\partial t}{\mathscr N}(E,t)=%
-\frac{\partial}{\partial E}\left[\frac{\ud E}{\ud t}{\mathscr N}(E,t)\right]%
-\frac{{\mathscr N(E,t)}}{t_{\rm esc,d}}+Q(E)\,,
\end{equation}
where $Q(E)$ is the injection rate at the shock front (see Eq.~\ref{injrate})
and $t_{\rm esc,d}$ accounts for downstream losses due to both advection,
which is produced by the downstream flow carrying the scattering centres of CRs,
and diffusion.
It can be written as
\be\label{tescd}
\frac{1}{t_{\rm esc,d}}=\frac{1}{t_{\rm adv}}+\frac{1}{t_{\rm diff}}\,.
\ee
The advection time is given by
\be\label{tadv}
t_{\rm adv}=\frac{R_{\rm HS}}{v_{\rm fl,d}}=\frac{rR_{\rm HS}}{(r-1)U}\,,
\ee
with $v_{\rm fl,d}$ the downstream flow velocity in the observer reference frame and
$R_{\rm HS}$ the radius of the hot spot region.
The downstream diffusion time reads
\be\label{tdiff}
t_{\rm diff} = \frac{R_{\rm HS}^{2}}{6\kappa_{\rm HS}}\,,
\ee
where the factor 6 accounts for three-dimensional diffusion.
The diffusion coefficient in the hot spot region, $\kappa_{\rm HS}$, reads
\be
\kappa_{\rm HS}=\varkappa_{\rm HS}\kappa_{\rm gal}\,,
\ee
which we assume is proportional to the local galactic diffusion coefficient, $\kappa_{\rm gal}$,
since we suppose that the turbulence self-generated at the shock surface has damped at large distances from the shock front. 
From radio galactic emission observations and
secondary-to-primary CR ratios, 
$\kappa_{\rm gal}$ deduced for the propagation in the local ISM reads 
\be\label{kappagal}
\kappa_{\rm gal}=4\times10^{28}\left(\frac{E}{3~\mathrm{GeV}}\right)^{0.5}~\mathrm{cm^{2}~s^{-1}}
\ee
for $E\ge3$~GeV (Berezinskii et al.~\cite{bb90}). For lower energies, $\kappa_{\rm gal}$ is not well constrained and we assume
$\kappa_{\rm gal}=4\times10^{28}~\mathrm{cm^{2}~s^{-1}}$.
Some recent estimates using Voyager~1 data (Herbst et al.~\cite{hh12}) give $\kappa_{\rm gal}\simeq10^{26}-10^{27}~\mathrm{cm^{2}~s^{-1}}$,
so we adopt $\varkappa_{\rm HS}=0.01-1$ to account for possible turbulence enhancement with respect to the local value.

The energy loss per unit time is described by
\be\label{dotE}
\left|\frac{\ud E}{\ud t}\right|=\frac{E}{t_{\rm ad}}%
+3\beta\left(\frac{n_{\rm H}}{10^{6}~\mathrm{cm^{-3}}}\right)\left[\frac{L(E)}{10^{-25}~\mathrm{GeV~cm^{2}}}\right]%
~\mathrm{GeV~s^{-1}}\,,
\ee
where $L(E)$ is the energy loss function described in Sect.~\ref{colllosses}. The adiabatic time, $t_{\rm ad}$, accounts
for the 
fact that behind the bow shock
there is a re-expansion of the flow so that CRs adiabatically lose energy; it is given by
\be\label{tad}
t_{\rm ad}=\psi\frac{r}{3(r-1)}\frac{R_{\rm HS}}{U}\,,
\ee
where  
$\psi$ is equal to 1.5 or 3 for non-relativistic or relativistic particles, respectively
(see e.g. Lerche \& Schlickeiser~\cite{ls82}). 

The injection rate at the shock front, $Q(E)$, which we assume to be time-independent, reads
\be\label{injrate}
Q(E)=\frac{{\mathscr N}_{\rm sh}(E)}{t_{\rm esc,a}}\,,
\ee
where ${\mathscr N}_{\rm sh}(E)$ is given by Eq.~(\ref{Nsh})
and $t_{\rm esc,a}$ is the escaping time from the acceleration zone. Following Moraitis \& Mastichiadis~(\cite{mm07}), 
we can write
\be\label{tesca}
t_{\rm esc,a}=\frac{r{\mathscr L}}{U}\,,
\ee
where ${\mathscr L}=k_{\rm u}^{-\alpha}\kappa_{\rm B}U^{-1}[1+r(k_{\rm d}/k_{\rm u})^{\alpha}]$ is the size of the region around the shock where the acceleration takes place. 
For electrons, Eq.~(\ref{tesca}) should include a term in the denominator for radiative losses, but it is negligible for energies lower than
TeV and a magnetic field strength lower than 1~mG.

\begin{figure}[!ht]
\begin{center}
\resizebox{\hsize}{!}{\includegraphics{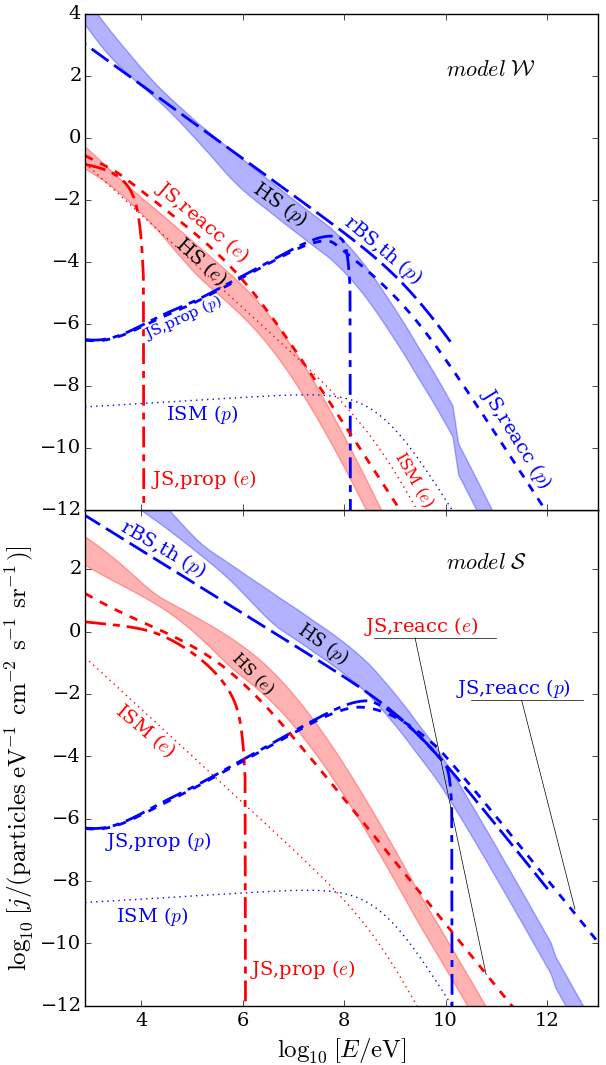}}
\caption{Spectra of accelerated protons ({\em blue})
and secondary electrons ({\em red}) for model \mW\ and \mS\ ({\em upper} and {\em lower} panel, respectively),
and a rBS at $1.8\times10^{3}$~AU.
Both panels show 
the spectra of the CRs accelerated at the jet shock at 100~AU and propagated up to the rBS (JS,prop; {\em dash-dotted lines}),
the spectra after re-acceleration at the rBS (JS,reacc; {\em short-dashed lines}), 
the spectrum of the accelerated CR protons at the rBS (rBS,th; {\em long-dashed line}),
the spectra in the hot spot region (HS; {\em shaded regions}),
and the interstellar CR proton and electron spectra (ISM; {\em dotted lines}).
}
\label{spectrumatrBS_paper_ok}
\end{center}
\end{figure}

In general, we can assume particles to be in a steady state since the lifetime, given by
\be\label{tlife}
t_{\rm life}=\frac{D_{\rm rBS}}{v_{\rm jet}}\,,
\ee
where $D_{\rm rBS}$ is the distance of the reverse bow shock, is much longer than the escape time downstream 
(see Appendix~\ref{app:lifetime}) and this allows us to put $\partial\mathscr{N}(E,t)/\partial t=0$ in Eq.~(\ref{Nevolve}).
Ginzburg \& Syrovatskii~(\cite{gs64}) give the analytical solution of Eq.~(\ref{Nevolve}) in the steady-state case,
\be\label{Nprop} 
{\mathscr N}(E)=\left(\frac{\ud E}{\ud t}\right)^{-1}\int_{E}^{E_{\rm max}}Q(E^{\prime})\exp\left[-\frac{\tau(E,E^{\prime})}{t_{\rm esc,d}}\right]\ud E^{\prime}\,,
\ee
where
\be
\tau(E,E^{\prime})=\int_{E}^{E^{\prime}}\left(\frac{\ud {E}^{\prime\prime}}{\ud t}\right)^{-1}\ud E^{\prime\prime}\,.
\ee
Finally, the solution of Eq.~(\ref{Nprop}) allows us to derive the particle spectra in the hot spot region, $j_{\rm HS}$, which reads
\be
j_{\rm HS}(E)=\frac{v_{\rm dr}(E)\mathscr{N}(E)}{4\pi}\,,
\ee
where
\be
v_{\rm dr}=\left(\frac{\kappa_{\rm HS}}{6t_{\rm life}}\right)^{0.5}
\ee 
is the drift velocity of the particles in the turbulent hot spot region
(Skilling~\cite{s75}). The resulting CR spectra in the hot spot region, labelled ``HS'' in Fig.~\ref{spectrumatrBS_paper_ok},
are encompassed by a shaded region; the upper and lower limits are obtained by assuming $\varkappa_{\rm HS}$ equal to 1 and 0.01, respectively.

We note that Ptuskin et al.~(\cite{pm06}) compute different trends for $\kappa_{\rm gal}$ at low energies. 
Therefore we also calculate the emerging spectra in the hot spot region by considering 
an increasing and a decreasing diffusion coefficient at non-relativistic energies, namely accounting for
the plain diffusion model (hereafter PD), 
$\kappa_{\rm gal}\propto\beta^{-2}$, and the diffusive re-acceleration model (hereafter DR), 
$\kappa_{\rm gal}\propto\beta(\gamma^{2}-1)^{0.17}$,
respectively. The resulting CR spectra, $j_{\rm HS}$, both for CR protons and electrons at most increase or decrease
by one order of magnitude in the non-relativistic energy range, using the PD or the DR model, respectively
(we note that $v_{\rm dr}\propto\kappa_{\rm gal}^{0.5}$).

Finally, Fig.~\ref{spectrumatrBS_paper_ok} also displays the expected CR proton and electron spectra propagating into the hot spot region after the rBS
acceleration (see Sect.~\ref{twozonemodel}) and the interstellar CR spectra constrained by 
the most recent observations of the Alpha Magnetic Spectrometer (Aguilar et al.~\cite{aa14,aa15}), 
while at low energies we used the results of Stone et al.~(\cite{sc13}) based on Voyager observations
(see also Ivlev et al.~\cite{ip15}).
A comparison between the hot spot and the interstellar CR spectra confirms that effects such as synchrotron emission for a shock such as
model \mS\ (Sect.~\ref{dgtau})
and high ionisation rates (Sect.~\ref{l1157b1} and~\ref{omc2fir4}) 
cannot be due
to the interstellar CR flux, but could be explained by a source of accelerated particles in jet shocks.

\section{Cosmic-ray ionisation rate}\label{ionrates}
In the next subsections we examine the range of values of the ionisation rate expected inside a jet accounting for the different
effects described in Sects.~\ref{CRpressureeffects}, \ref{keffects}, and \ref{energylossregime} (diffusion coefficient, shock
CR pressure, and gyromotion, respectively). Then we investigate the impact on the heating of the protostellar disc due to these locally
accelerated CRs.

\subsection{Ionisation rate along the jet}
We use the emerging CR spectra of the jet shock-accelerated protons (models \mW\ and \mS) shown in Fig.~\ref{emergingspectra} 
to compute the CR ionisation rate, \zhh, which reads%
\be\label{zetaH2}
\Ezhh(N)=2\pi\sum_{k=p,e}\int j_{k}(E,N)\sigma_{k}^{\rm ion}(E)\ud E\,,
\ee
where $j_{k}$ is the spectrum of the accelerated CR protons or secondary electrons (see Eq.~\ref{jCR} and Sect.~\ref{energylossregime}).
We apply the modelling described in Padovani et al.~(\cite{pgg09}) to study the variation of \zhh\ with increasing column density, i.e.
while CR protons propagate inside the jet.
With respect to Padovani et al.~(\cite{pgg09}), 
cross sections, $\sigma_{k}^{\rm ion}$, were modified to include the effect of 
relativistic protons and electrons (see Krause et al.~\cite{km15}). 
This does not invalidate the previous results since 
the spectra used in Padovani et al.~(\cite{pgg09}) have a 
negligible high-energy component, so that any former conclusion remains 
accurate. 
The resulting ionisation rates at the shock surface
show values between $3\times10^{-10}$~s$^{-1}$ (model \mW) and $5\times10^{-9}$~s$^{-1}$ 
(model \mS).

\begin{figure}[!ht]
\begin{center}
\resizebox{\hsize}{!}{\includegraphics{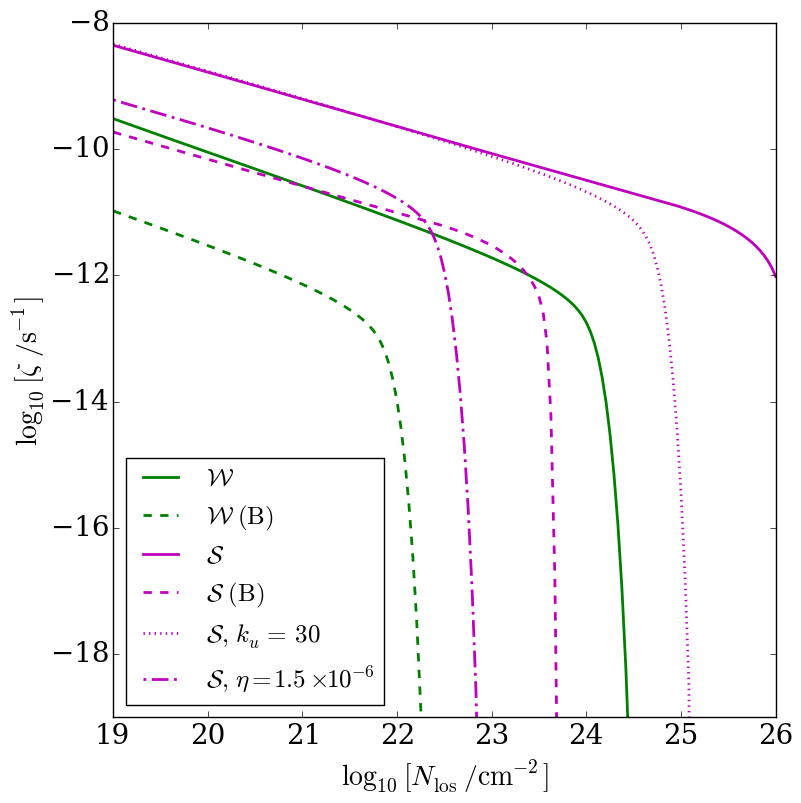}}
\caption{Ionisation rate at the shock surface in a jet
as a function of the molecular hydrogen column density for models \mW\ and \mS\ described in the main text
in the case of a parallel shock.
{\em Solid} and {\em dashed lines} show the values of \zhh\ neglecting and accounting for gyromotion effects, respectively.
The {\em dotted line} displays \zhh\ for model \mS\ when the upstream diffusion coefficient is 30 times higher than the Bohm
coefficient, without including any further attenuation due to gyromotion. The {\em dash-dotted} line shows \zhh\ for model \mS\
with $\eta$ reduced by a factor of about ten with respect to the value in Table~\ref{paramf0}.
}
\label{zetaWS}
\end{center}
\end{figure}

Because of the strong toroidal magnetic field configuration expected in jets (Blandford \& Payne~\cite{bp82}), we assume that the transverse
diffusion is negligible so that the accelerated CRs are confined to the jet.
Figure~\ref{zetaWS} displays the total ionisation rate (primary protons plus secondary electrons) 
for increasing column density along the line of sight for the two models
\mW\ and \mS, hereafter labelled  \zW\ and \zS, respectively. 

Without including gyromotion effects, \zW\ is about one order of magnitude lower than \zS\ at low column densities, 
\mbox{$N\lesssim10^{22}~\mathrm{cm^{-2}}$}. Then, around 
$N$=10$^{23}$~cm$^{-2}$ \zW\ decreases ever more  rapidly until CRs are completely thermalised around
$3\times10^{24}$~cm$^{-2}$. Conversely, model \mS\ is not strongly attenuated, not even at 
$N=10^{26}$~cm$^{-2}$, because of the larger reservoir of high-energy CRs that gives an efficient ionisation at higher 
column densities (see also Sect.~\ref{energylossregime}).
If we account for the impact of gyromotion, both \zW\ and \zS\ decrease by about one order of magnitude
at low column densities, but then CRs are more rapidly thermalised at about $2\times10^{22}$~cm$^{-2}$ and about
$4\times10^{23}$~cm$^{-2}$ for model \mW\ and \mS, respectively. 

Figure~\ref{zetaWS} also shows the dependence of \zhh\ on column density with an upstream diffusion coefficient 
$\kappa_{\rm u}=30\,\kappa_{\rm B}$. As explained in Sect.~\ref{keffects}
(see also Figs.~\ref{nxT4ku30} and \ref{emax_vs_ku}), $E_{\rm max}$ decreases for increasing values of $\kappa_{\rm u}$,
and accelerated CRs are thermalised at a lower column density ($N\simeq10^{25}$~cm$^{-2}$).

We finally evaluate
the trend of the ionisation rate for model \mS\ assuming a shock efficiency $\eta=1.5\times10^{-6}$ (Eq.~\ref{shockeff}), 
which is about one order of magnitude 
lower than the value used in the previous sections. 
The drop in \zhh\ is due to fact that when $\eta$ decreases, then $E_{\rm max}$ is fixed by $E_{\rm damp}$ (Eq.~\ref{Edamp}).
In this case the accelerated CR protons are thermalised at 
a column density more than three orders of magnitude lower ($N\simeq7\times10^{22}~\mathrm{cm^{-2}}$)
than in the case of $\eta=10^{-5}$. 

Even if the knowledge of some parameters constraining our modelling is still missing (e.g. shock efficiency, 
magnetic field strength and configuration, turbulent
magnetic field component), we can imagine that the accelerated CR flux and the corresponding ionisation rate
could be even markedly modified by the mechanisms discussed in this paper.
Whatever the involved processes, the ionisation rate due to the shock particle acceleration may be higher than the
typical values of $10^{-17}-10^{-18}$~s$^{-1}$ estimated for dense cores and protostellar discs due to the interstellar cosmic-ray flux,
at least in a region close to the acceleration site.

\subsection{Ionisation and heating rates in a protostellar disc}\label{heating}
\begin{figure}[!ht]
\begin{center}
\resizebox{\hsize}{!}{\includegraphics{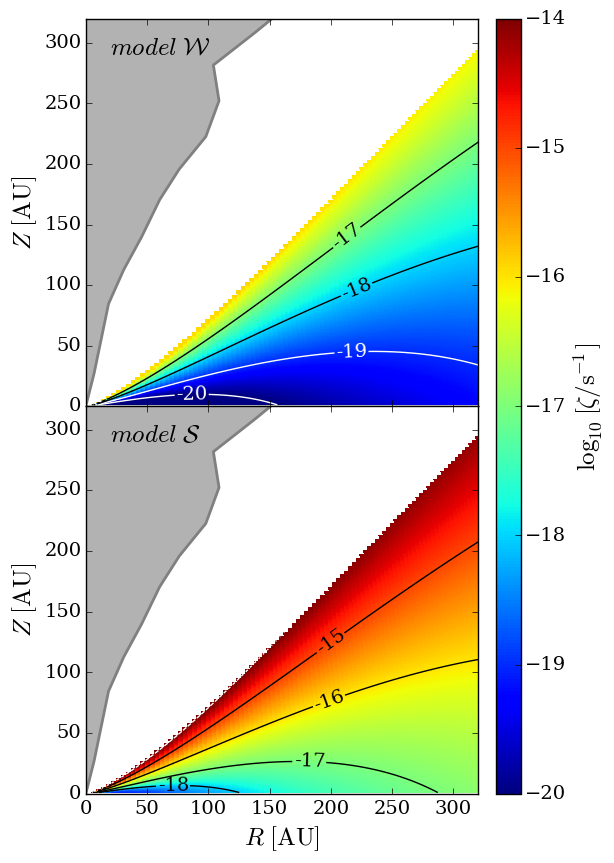}}
\caption{CR ionisation rate profile in a protostellar disc according to model ${\cal W}$ ({\em upper panel}) and ${\cal S}$
({\em lower panel}). Black and white solid lines show the iso-ionisation rate contours. The {\em grey shaded area} shows the jet profile according
to the jet opening angle estimated by Dougados et al.~(\cite{dc00}).
}
\label{zetainenvelope_OK}
\end{center}
\end{figure}

In order to quantify the variation of the ionisation degree in a protostellar disc due to accelerated CRs propagating from the hot spot region, 
we use the two-dimensional disc density profile described in Andrews et al.~(\cite{aw11}) and Cleeves et al.~(\cite{ca13}).
We propagate the hot spot CR proton and electron spectra computed for
both models ${\cal W}$ and ${\cal S}$ (see Fig.~\ref{spectrumatrBS_paper_ok}) accounting for
a dilution factor $d^{-2}$, where $d$ is the distance from the hot spot, and we consider the distance of the
hot spot region equal to $1.8\times10^{3}$~AU as in DG Tau (Sect.~\ref{twozonemodel}).
Figure~\ref{zetainenvelope_OK} shows the expected ionisation rate in the disc: while for model ${\cal W}$ 
the ionisation rate is comparable to that due to the interstellar CR flux, for model ${\cal S}$ the value of $\zeta$ increases up to about $10^{-14}$~s$^{-1}$
in the upper layers of the protostellar disc.

In order to check that $\zeta\approx10^{-14}$~s$^{-1}$ does not give a gas temperature, $T_{\rm g}$, that is higher than the observed values,
we follow the approach outlined by Goldsmith~(\cite{g01}) and Galli et al.~(\cite{gw02})
used to compute $T_{\rm g}$ by balancing the heating by the locally accelerated CRs
and the cooling due to molecular
and atomic transitions and collisions with dust grains. In the approximation where the dust temperature, $T_{\rm d}$, is independent of
interaction with the gas, the thermal balance equation is given by
\be
\Gamma_{\rm CR}=\Lambda_{\rm gd}+\Lambda_{\rm g}\,,
\ee
where $\Lambda_{\rm gd}$ is the gas-dust energy transfer rate, given by Burke \&
Hollenbach~(\cite{bh83}) and is given by
\be
\Lambda_{\rm gd}=2\times10^{-33}\left(\frac{n_{\rm H}}{\mathrm{cm^{-3}}}\right)^{2}\left(\frac{T_{\rm g}-T_{\rm d}}{\mathrm{K}}\right)%
\left(\frac{T_{\rm g}}{10~\mathrm{K}}\right)^{1/2}~\mathrm{erg~cm^{-3}~s^{-1}}\,,
\ee
while $\Lambda_{\rm g}$ is the gas cooling rate by molecular and atomic transitions, given by Goldsmith~(\cite{g01})
\be
\Lambda_{\rm g}=\alpha_{\rm g}\left(\frac{T_{\rm g}}{10~\mathrm{K}}\right)^{\beta_{\rm g}}~\mathrm{erg~cm^{-3}~s^{-1}}\,,
\ee
where $\alpha_{\rm g}$ and $\beta_{\rm g}$ are parameters that depend on the total hydrogen density and the molecular depletion
factor, $f_{\rm d}$.
We adopt a dependence of $f_{\rm d}$ on $n_{\rm H}$ given by $f_{\rm d}=\exp(n_{\rm H}/n_{\rm dep})$ or
$f_{\rm d}=f_{\rm d,max}$ for $n_{\rm H}\le n_{\rm dep}\log(f_{\rm d,max})$ or $n_{\rm H}>n_{\rm dep}\log(f_{\rm d,max})$, 
respectively. The critical density for CO depletion is taken to be
$n_{\rm dep}=5.5\times10^{4}$~cm$^{-3}$ and the maximum depletion factor is $f_{\rm d,max}=100$.
The CR heating rate reads
\be
\Gamma_{\rm CR}=\left(\frac{n_{\rm H}}{\mathrm{cm^{-3}}}\right)%
\left(\frac{\zeta}{\mathrm{s^{-1}}}\right)%
\left(\frac{E_{\rm h}}{\mathrm{erg}}\right)\,,
\ee
where $E_{\rm h}$ is the mean heat input per ionisation (Glassgold et al.~\cite{ggp12}).
Neglecting the UV heating by the interstellar radiation field, we can estimate the net
effect of locally accelerated CRs on the gas temperature.
We consider two positions in the disc upper layers at a radius $R=50$~AU ($n_{\rm H}=7\times10^{5}$~cm$^{-3}$, $T_{\rm d}=100$~K) 
and $R=200$~AU ($n_{\rm H}=8\times10^{4}$~cm$^{-3}$, $T_{\rm d}=30$~K), e.g. Cleeves et al.~(\cite{ca13}). 
Assuming that the column density passed through the accelerated CRs coming from the hot spot is $10^{19}$~cm$^{-2}$ and
$\zeta=10^{-14}$~s$^{-1}$, we find $T_{\rm g}=130$~K and 108~K at $R=50$~AU and 200~AU, respectively.
These values of $T_{\rm g}$ are comparable with those estimated by Cleeves~(\cite{ca13}).

\section{Comparison with observations}\label{observations}
In the following subsections we describe a number of examples where we adopt our modelling as a theoretical support for observations.

\subsection{Synchrotron emission in DG Tau}\label{dgtau}
Ainsworth et al.~(\cite{as14}) detected synchrotron emission towards the low-mass T Tauri star DG~Tau through observations at low frequencies
(325 and 610~MHz) speculating that this could be due to relativistic electrons accelerated in the interaction between the jet and the ambient
medium. This emission is associated with a bow shock at a distance of about 1800~AU, 13$^{\prime\prime}$ 
from the central source, which is moving at about 100~km~s$^{-1}$ 
(knot C in Eisl\"offel \& Mundt~\cite{em98}). The minimum magnetic field strength and particle energy that can explain the synchrotron 
emission is $B_{\rm min}\approx110~\mu$G and $E_{\rm tot}\approx2.5\times10^{52}$~eV (Ainsworth et al.~\cite{as14}).

The jet structure of DG Tau is well studied down to subarcsecond scales (Maurri et al.~\cite{mb14}) where other knots and inner bow shocks
were discovered in addition to the arcsecond-scale knots (Eisl\"offel \& Mundt~\cite{em98}). Moreover,
McGroarty et al.~(\cite{mp09}) and Oh et al.~(\cite{op15}) computed kinematic and physical properties along the jet.
Even if we are aware of this complex structure, we show that the synchrotron emission seen towards DG~Tau
can be explained by the acceleration of secondary CR electrons in the jet-hot spot system.
According to our model
constraints (Sect.~\ref{Sect2}),  it is not possible to have an efficient acceleration of thermal electrons, although secondary CR electrons
produced in a previous shock can be re-accelerated (see Sect.~\ref{reaccatrBS}). Through this mechanism 
CR electrons gain a noticeable boost
at relativistic energies at the rBS surface.

For this reason, we suppose that a first acceleration takes place at the shock surface of an inner knot (knot B in Eisl\"offel \& Mundt~\cite{em98}),
computing the emerging CR spectrum according to Sect.~\ref{spectra}, and
we follow the propagation of the accelerated CR protons and secondary electrons up to knot C (as in Sect.~\ref{energylossregime})
accounting for gyromotion effects. 
Then we consider re-acceleration at the rBS (Sect.~\ref{reaccatrBS})
and diffusion in the hot spot region (Sect.~\ref{twozonemodel}).
Finally, following Longair~(\cite{l11}), assuming an isotropic distribution of re-accelerated secondary electrons in the hot spot region
with Lorentz factors in the range $[\gamma_{\rm min},\gamma_{\rm max}]$, we estimate their synchrotron emissivity, 
$\varepsilon_{\nu}$, which is given by
\be
\varepsilon_{\nu}=\frac{1}{4\pi}\int_{\gamma_{\rm min}}^{\gamma_{\rm max}}n_{e}(\gamma)P_{\rm S}(\nu,\gamma)\ud\gamma\,,
\ee
where $P_{\rm S}(\nu,\gamma)$ 
is the power emitted at frequency $\nu$ by a single electron with Lorentz factor $\gamma$ averaged over all possible
directions, which reads
\be
P_{\rm S}(\nu,\gamma)=\frac{\sqrt{3}e^{3}\langle B_{\perp}\rangle}{m_{e}c^{2}}F\left(\frac{\nu}{\nu_{c}}\right)\,,
\ee
with
\be
F(x)=x\int_{x}^{\infty}K_{5/3}(\xi)\ud\xi\,,
\ee
where $x=\nu/\nu_{c}$, $K_{5/3}$
is the modified Bessel function of order 5/3, and
\be
\nu_{c}=\frac{3}{2}\frac{\gamma^{2}e\langle B_{\perp}\rangle}{2\pi m_{e}c}\,,
\ee
where $\langle B_{\perp}\rangle$ is the average value of the perpendicular component of the magnetic field, which is equal to $\pi B/4$  
for an isotropic electron population.
The electron density per unit volume, $n_{\rm e}(\gamma)$, is given by 
\be
n_{\rm e}(\gamma)=\frac{4\pi j(E)}{v(E)}\frac{\ud E}{\ud\gamma}\,.
\ee
The synchrotron emissivity has to be converted into spectral energy flux density, $S_{\nu}$, so as to compare the model to 
Giant Metrewave Radio Telescope (GMRT) and Expanded Very Large Array (EVLA) observations
of DG~Tau (Ainsworth et al.~\cite{as14} and Lynch et al.~\cite{lm13}, respectively). Assuming a Gaussian beam profile, $S_{\nu}$ reads
\be
S_{\nu}=\frac{\pi}{4\ln2}I_{\nu}\theta_{\rm FWHM}^{2}\,,
\ee
where $\theta_{\rm FWHM}$ is the synthesised beam size in radians.
The specific intensity, $I_{\nu}$, is given by
\be
I_{\nu}=\frac{\varepsilon_{\nu}}{\kappa_{\nu}}\left(1-e^{-\tau_{\nu}}\right)\,,
\ee
where $\tau_{\nu}=R\kappa_{\nu}$ is the optical depth and $R$ is the radius of the emitting region.
Finally, the specific absorption coefficient, $\kappa_{\nu}$, reads
\be
\kappa_{\nu}=\frac{1}{8\pi m_{e}\nu^{2}}\int_{\gamma_{\rm min}}^{\gamma_{\rm max}}\gamma^{2}P_{\rm S}(\nu,\gamma)\frac{\ud}{\ud\gamma}%
\left[\frac{n_{\rm e}(\gamma)}{\gamma^{2}}\right]\ud\gamma\,.
\ee

\begin{figure}[!ht]
\begin{center}
\resizebox{\hsize}{!}{\includegraphics{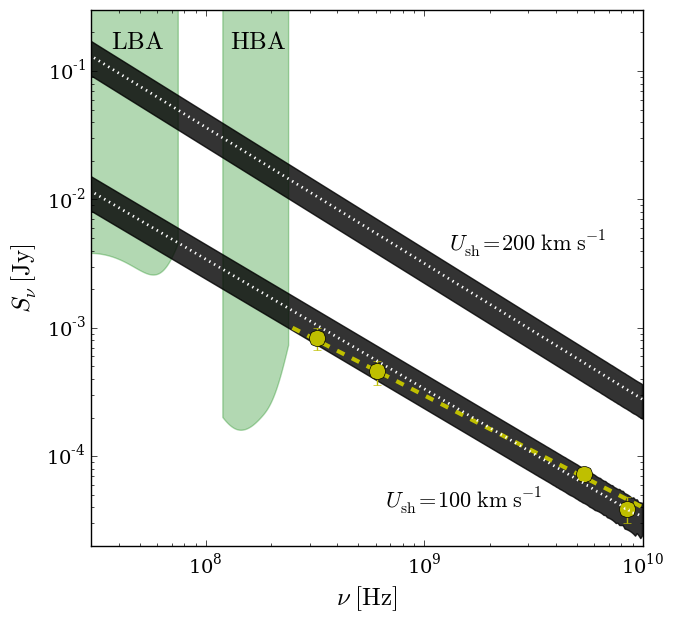}}
\caption{Spectral energy flux density as a function of the frequency.
GMRT and EVLA observations ({\em yellow solid circles}) from Ainsworth et al.~(\cite{as14}) and Lynch et al.~(\cite{lm13}), respectively,
are shown together with their fit ({\em yellow dashed line}) predicting a synchrotron spectral index of $-0.89\pm0.07$
(Ainsworth et al.~\cite{as14}). 
The two {\em black shaded} regions show the result of our modelling using two different shock velocities with respect to the upstream flow
($U=100$~km~s$^{-1}$ and 200~km~s$^{-1}$) and their widths refer to an assumed error of 30\% on the value of the hot spot
radius, the central value ($R_{\rm HS}=1300$~AU, Ainsworth et al.~\cite{as14}) pinpointed by the {\em white dotted lines}.
The {\em green shaded} areas show the two LOFAR bands (LBA=low band antenna; HBA = high band antenna), and their lower boundary in $S_{\nu}$
corresponds to the sensitivity limit using its most extended configuration (van Haarlem et al.~\cite{vhw13}). 
}
\label{ainsworth}
\end{center}
\end{figure}

Relying on the above equations, we compute the expected synchrotron emission spectrum.
We assume a constant magnetic field strength of $300~\mu$G 
for both knots B and C, $R$ of the order of 1300~AU (Ainsworth et al.~\cite{as14}), and 
we compute the expected synchrotron emission for
two values of the shock velocity with respect to the upstream flow 
($U=100$~km~s$^{-1}$ and 200~km~s$^{-1}$) because of the uncertainty in its value. We also adopt an 
error of 15\% in the emitting region radius.
As shown by Fig.~\ref{ainsworth}, 
we find a synchrotron spectral index of $-1.01$, which is close to the value inferred from observations ($-0.89\pm0.07$, 
Ainsworth et al.~\cite{as14}), and we are able to explain the observations with $U=100$~km~s$^{-1}$. 
It is important to note that the synchrotron emission would be inefficient if secondary CR electrons were not re-accelerated in
subsequent shocks: this is the key process to accelerate CR electrons in the synchrotron energy domain (see Fig.~\ref{spectrumatrBS_paper_ok}).
Finally, we also show in Fig.~\ref{ainsworth} the frequency range that was  observed by LOFAR, which can give a further constraint to our 
model.

\subsection{High ionisation rate in L1157-B1}\label{l1157b1}
Another case of remarkably high level of ionisation was measured by Podio et al.~(\cite{pl14}) in the bow shock of L1157 
known as B1. They found that 
the observed 
abundances of the HCO$^{+}$ and N$_{2}$H$^{+}$ ions, which are usually employed such as probes of the ionisation rate, can be
simultaneously reproduced only by assuming $\zeta=3\times10^{-16}$~s$^{-1}$. 
The youngest knot, termed B0, lies at about $1.2\times10^{4}$~AU, while B1 is at $1.7\times10^{4}$~AU with a hot spot cavity radius
of about $1.2\times10^{3}$~AU (Lefloch et al.~\cite{lc12}), assuming a source distance of 250~pc (Looney et al.~\cite{lt07}).
The jet velocity is of the order of
$100$~km~s$^{-1}$ with shock velocities between 20 and 40~km~s$^{-1}$ (Bachiller et al.~\cite{bp01}; Tafalla et al.~\cite{tb15}).
The total hydrogen density is of the order of $10^{5}-10^{6}$~cm$^{-3}$ (e.g. G\'omez-Ruiz et al.~\cite{gc15}). 
Podio et al.~\cite{pl14} traced the cold gas with temperatures of $60-200$~K and
there are hints of a warm gas component with a temperature of at least $10^{3}$~K that can  
explain the water lines (Busquet et al.~\cite{bl14}). 

Our modelling can explain the high ionisation rate in B1 if we assume that thermal protons are accelerated in B0
and, once they reach B1, that they undergo a further acceleration without any additional thermal proton acceleration.
This situation can take place supposing that 1) the shock velocity with respect to the upstream flow
and/or the shock efficiency is  too low 
at the bow-shock surface
B1 or 2) the upstream diffusion coefficient is too large to have an efficient acceleration of thermal particles.  

We follow the steps described in Sects.~\ref{spectra} and 
\ref{energylossregime}
(calculation of the emerging CR spectrum at B0, 
its propagation up to B1, 
its re-acceleration at B1, 
and the hot spot region diffusion),
assuming $U=60$~km~s$^{-1}$ and $U=40$~km~s$^{-1}$ in B0 and B1, respectively, 
$N_{\rm eff}/N_{\rm los}=2\times10^{2}$ (see Sect.~\ref{energylossregime}),
$B\lesssim100~\mu$G, $x=0.2-0.4$, $T=10^{3}$~K, $n_{\rm H}=10^{5}$~cm$^{-3}$, $k_{\rm u}=1$, and $\eta=5\times10^{-6}$ in both B0 and B1.
We obtain proton and secondary electron spectra leading to 
a total ionisation rate $\Ezhh=6.1\times10^{-16}$~s$^{-1}$, consistent within a factor of about 2 with the value estimated from observations.

The values of $U$, $B$, $T$, $x$, $n_{\rm H}$, $k_{\rm u}$, and $\eta$ 
can vary along the shock surfaces B0 and B1, which is why our result has to be interpreted as a proof
of concept. Further observations could give better constraints on the above parameters to test the validity of our hypothesis.

\begin{figure}[!ht]
\begin{center}
\resizebox{\hsize}{!}{\includegraphics{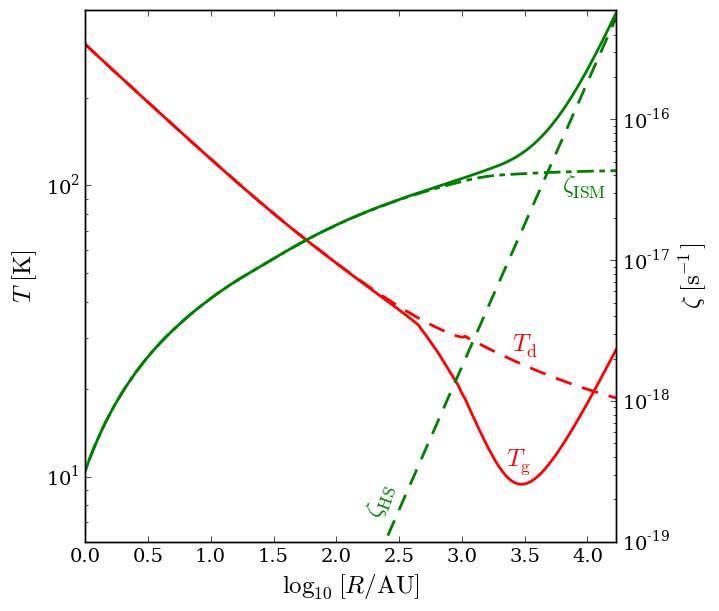}}
\caption{Dust and gas temperature ({\em solid} and {\em dashed red lines}, respectively) as a function of the distance from the protostar L1157.
The plot also shows the ionisation rate of interstellar CRs ({\em dash-dotted green line}), of CRs coming from the hot spot 
({\em dashed green line}), and the total value ({\em solid green line}).}
\label{L1157_temperature}
\end{center}
\end{figure}

Figure~\ref{L1157_temperature} shows the comparison between the ionisation rate due to the local CRs accelerated in the hot spot,
$\zeta_{\rm HS}$, and the value due to the interstellar 
CRs assuming a spectrum similar to that from Voyager~1, $\zeta_{\rm ISM}$ (Stone et al.~\cite{sc13}; Ivlev et al.~\cite{ip15}).
We conclude that the high value of \zhh\ observed in L1157 and explained by our modelling may not be due to interstellar
CRs. In fact,
at the hot spot ($R=1.7\times10^{4}$~AU), $\zeta_{\rm HS}$ is about a factor of 10 higher than $\zeta_{\rm ISM}$.
Then, entering the envelope towards the protostar, the contribution of the hot spot CR flux to the total ionisation rate becomes negligible 
at $R\lesssim5\times10^{3}$~AU because of the geometric dilution factor, $d^{-2}$, where $d$ is the distance from the hot spot.

Following Sect.~\ref{heating}, we compute the gas temperature in the envelope of L1157 only accounting for the heating due to both
interstellar and locally accelerated CRs. The dust temperature profile is given by $T(r)=300(R/\mathrm{AU})^{-0.41}$~K (Chiang et al.~\cite{cl10,cl12}).
Neglecting the UV heating by the interstellar radiation field, for $R\lesssim300$~AU gas and dust are coupled, while for larger radii, 
at first $T_{\rm g}$ decreases because the heating by the interstellar CRs is too weak, then CRs at the hot spot cause
a slight increase in $T_{\rm g}$ with respect to $T_{\rm d}$, up to 30~K.

\subsection{High ionisation rate in OMC-2 FIR 4}\label{omc2fir4}
Ceccarelli et al.~(\cite{cd14}) observed one of the closest known intermediate-mass protostars in Orion, OMC-2 FIR~4,
via {\em Herschel} observations of HCO$^{+}$ and N$_{2}$H$^{+}$. The abundance of these molecular ions were used to estimate the ionisation rate 
which reaches
values of the order of $1.5\times10^{-12}$ and $4\times10^{-14}$~s$^{-1}$ at 1600~AU and 3700~AU from the source centre, respectively.
Actually, the structure of this protostar is highly complex since it contains a cluster of a few embedded intermediate- 
and low-mass protostars (Shimajiri et al.~\cite{st08}; L\'opez-Sepulcre et al.~\cite{lt13}).
It is also worth noting that Shimajiri et al.~(\cite{st08}) found the presence of a shock region 
produced by the interaction of an external bipolar outflow driven by the nearby OMC-2 FIR~3 region, lying to the north-east of OMC-2 FIR~4
(see also L\'opez-Sepulcre et al.~\cite{lt13}). This shock
appears to be responsible for the high degree of fragmentation observed in OMC-2 FIR~4.

Thus far no jet activity has been observed in OMC-2 FIR~4. Nevertheless, in order to justify the extremely high values of
\zhh,  the presence of a mechanism able to accelerate particles inside the source must be  postulated since at a radius of the order 
of thousands of AU the interstellar CR flux is strongly attenuated (Padovani et al.~\cite{phg13}).
Assuming that no jet is present and that the accretion on the protostellar surface is still spherical, we can recover 
an efficient proton acceleration at the protostellar surface (Sect.~\ref{stellarsurfaceacc}).
For instance, following Sect.~\ref{Sect2} we compute the emerging proton spectrum by considering $U=260$~km~s$^{-1}$,
$T=9.4\times10^{5}$~K, $n_{\rm H}=1.9\times10^{12}$~cm$^{-3}$, $B=5$~G, $x=0.3$, $k_{\rm u}=1$, and $\eta=10^{-5}$ at $R_{\rm sh}=2\times10^{-2}$~AU
as in the model by Masunaga \& Inutsuka~(\cite{mi00}). The maximum energy of the accelerated CRs is $E_{\rm max}=11.4$~GeV. 
Then, neglecting magnetic turbulence (see Sect.~\ref{energylossregime}),
we propagate the proton and secondary electron fluxes up to the two positions where the ionisation rate in OMC-2 FIR~4 was 
estimated, 1600~AU and 3700~AU. Combining the density profiles in Masunaga \& Inutuska~(\cite{mi00}) and Crimier et al.~(\cite{cc09}),
we find that the accelerated particles go through a column density of about $1.8\times10^{24}$~cm$^{-2}$.
Accounting for a geometrical dilution factor of 
$(R_{\rm sh}/R)^{2}$, with $R=1600$~AU and 3700~AU, we obtain  
$\Ezhh=3\times10^{-15}$~s$^{-1}$ and $6\times10^{-16}$~s$^{-1}$, respectively. However, 
including gyromotion effects would increase the effective column density (Sect.~\ref{energylossregime}),
causing a further reduction in the ionisation rate.

It is important to emphasise that the above result is obtained by neglecting any diffusion process. Since it is difficult to describe
turbulence in the envelope, we briefly discuss the effects of diffusion here. If the accelerated CRs undergo diffusion
during propagation from the shock surface, the CR energy distribution function at a distance $R$ from the shock and at a time $t$ 
in the case of continuous CR injection from a point source reads
\be
{\mathscr N}(R,t,E) = \frac{Q(E)}{4\pi\kappa(E)R}\mathrm{erfc}(g)\,,
\ee
where $\kappa(E)$ is the diffusion coefficient, $g=R/(R_{\rm diff}\sqrt{2})$, and $R_{\rm diff}=\sqrt{t\kappa(E)}$ is 
the diffusion length (Aharonian~\cite{a04}).
If $R\gg R_{\rm diff}$, then 
$\mathrm{erfc}(g)\rightarrow 1/(g\sqrt{\pi}) \exp(-g^2)$ and the propagated spectrum is attenuated as $R^{-2}\exp[-R^{2}/(2R_{\rm diff}^{2})]$,
i.e. much faster than in the free-streaming case.
In contrast, if $R\ll R_{\rm diff}$, then $\mathrm{erfc}(g)\rightarrow1$ and the propagated spectrum is attenuated as $R^{-1}$,
as opposed to $R^{-2}$ in the free-streaming case, and $\zeta$ would be
much higher: $\zeta=3\times10^{-10}$~s$^{-1}$ and $\zeta=1\times10^{-10}$~s$^{-1}$ at $R=1600$~AU and $R=3700$~AU, respectively.

\begin{figure}[!ht]
\begin{center}
\resizebox{\hsize}{!}{\includegraphics{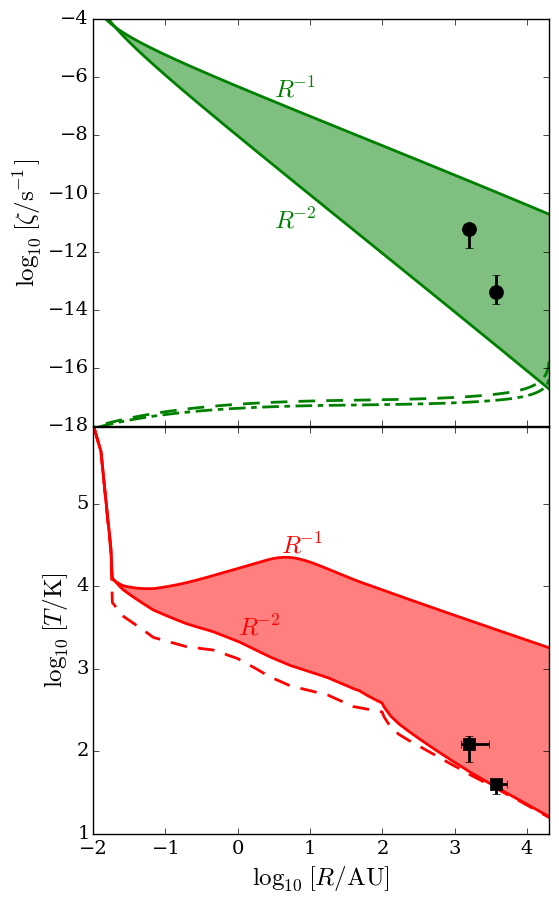}}
\caption{Ionisation rate (upper panel) and temperatures of gas and dust (lower panel) as a function of the distance from the protostar OMC-2 FIR~4. 
Observational estimates of $\zeta$ and $T_{\rm g}$ 
({\em black solid circles} and {\em squares}, respectively; Ceccarelli et al.~\cite{cd14}) 
are compared to the results from the modelling described in Sect.~\ref{omc2fir4} ({\em green} and {\em red solid lines}). 
The {\em green} and {\em red shaded areas} encompass the range of $\zeta$ and $T_{\rm g}$ by assuming 
a dilution factor $R^{-1}$ (purely diffusive propagation) and $R^{-2}$ (free-streaming propagation). 
The {\em green dash-dotted} and {\em dashed}
lines show the interstellar CR ionisation rate 
assuming a spectrum similar to that from Voyager 1
(Stone et al.~\cite{sc13}) 
and an enhanced CR proton flux (see model $\mathscr{H}$ in Ivlev et al.~\cite{ip15}).
The {\em red dashed line} shows the trend of the dust temperature (Masunaga \& Inutsuka~\cite{mi00}; Crimier et al.~\cite{cc09}).
}
\label{omc2fir4ionisation}
\end{center}
\end{figure}

As shown in Fig.~\ref{omc2fir4ionisation}, the values of $\zeta$ computed from observations lie between the two extreme cases of
free streaming and pure diffusion with $R\ll R_{\rm diff}$.\footnote{Hereafter
we refer to the purely diffusive case with $R\ll R_{\rm diff}$ as pure diffusion.} 
However, after performing a check on gas temperature (as outlined in Sect.~\ref{heating}),
the attenuation of the spectrum with a factor $R^{-1}$ corresponds to  higher values of $T_{\rm g}$ 
than   those determined by Ceccarelli et al.~(\cite{cd14}) from a high-velocity gradient analysis. 
We can conclude that the propagation mechanism is probably neither purely diffusive nor free streaming.
A better knowledge of the magnetic
field configuration close to the protostar and of its turbulence degree is needed for a more careful description of CR propagation from
the protostellar surface to the envelope.

\subsection{$^{10}$Be enrichment in meteorites}\label{10be}
Ceccarelli et al.~(\cite{cd14}) argued that the accelerated proton flux causing the high ionisation rate could also be
responsible for the formation of short-lived radionuclei, such as $^{10}$Be, contained in the calcium-aluminium-rich inclusions (CAIs) of
carbonaceous meteorites. In fact, the measured abundance of $^{10}$Be in meteorites is higher  than that found in the ISM and this
should be due to spallation reactions between the accelerated CRs and the thermal gas that took place during 
the earliest phases of the protosolar nebula.

To test the consequences of our modelling, we use the emerging spectrum at the protostellar surface 
(Sect.~\ref{omc2fir4}) scaled at 1~AU, taking into account the geometric dilution of $R^{-1}$ or $R^{-2}$, appropriate for 
the purely diffusive or free-streaming case, respectively.
We calculate the fluence per unit time, $\mathscr{F}_{t}$, which reads
\be\label{eqflu}
\mathscr{F}_{t}(E_{\rm min})=2\pi\int_{E_{\rm min}}^{E_{\rm max}}j(E)\ud E\,,
\ee
where $E_{\rm min}\simeq50$~MeV is the energy threshold for the spallation reaction 
$p+\mathrm{^{16}O}\rightarrow\!^{10}\mathrm{Be}+\ldots$ (Gounelle et al.~\cite{gs06})
and $E_{\rm max}=11.4$~GeV (Sect.~\ref{omc2fir4}).
We find $\mathscr{F}_{t}=2\times10^{17}$ and $8\times10^{18}$~protons~cm$^{-2}$~yr$^{-1}$ at 1~AU, 
for the purely diffusive and free-streaming cases, respectively.
This means that
an irradiation time of a few tens of years 
can explain the values of the fluence derived by Gounelle et al.~(\cite{gc13}) equal to 
$10^{19}-10^{20}$~protons~cm$^{-2}$, 
in agreement with the estimates by Ceccarelli et 
al.~(\cite{cd14}).
This result is also consistent with the X-wind model predictions (Shu et al.~\cite{ss97}), according to which a proto-CAI of radius $R_{\rm CAI}$
spends a time $t\sim20(R_{\rm CAI}/\mathrm{1~cm})$~yr in the reconnection ring.
Nevertheless, it is important to note that the model by Gounelle et al.~(\cite{gc13}) predicts too much heating
(Tatischeff et al.~\cite{td14}).
If diffusion is present, the particle flux in Eq.~(\ref{eqflu}) is higher and the fluence computed by Gounelle et al.~(\cite{gc13}) would be
even more easily recovered.

\section{Conclusions}\label{conclusions}
In this paper we investigated the possibility of accelerating CRs within a protostellar source by means of shock processes.
Diffusive shock acceleration (DSA) is the main process leading to CR acceleration: a CR gains energy 
up to the relativistic domain by multiple
back-and-forth shock crossings. A number of conditions have to be satisfied in order for DSA to be effective, some of them related to 
magnetic fluctuations (determining pitch-angle scattering), others associated 
with cooling processes and also shock velocity, age, and geometry constraints. We focused our attention on the effectiveness of shocks in
accretion flows, in jets, and on protostellar surfaces. Our main conclusions are the following:
\begin{itemize}
\item[1.] At jet shocks, CR protons can be effectively extracted from the thermal plasma and accelerated up to relativistic energies, while
CR electrons can 
only reach energies of
about 300~MeV because of wave damping and energy losses.
\item[2.] In accretion flows, the ionisation fraction is too small and the shock velocity  too low, which prevents
any CR acceleration. Furthermore,  the high magnetic field strength implies sub-Alfv\'enic flow speeds.
\item[3.] On protostellar surfaces, shocks caused by impacting material
during the collapse phase are strong enough to boost CR protons, but not electrons.

\end{itemize}
The set of conditions that has to be fulfilled is highly non-linear: small variations in one or more parameters 
(magnetic field strength, ionisation fraction, total hydrogen density, temperature, flow velocity, upstream diffusion coefficient, and shock 
efficiency) can 
make the acceleration process inefficient. As a consequence, since a protostar is a highly dynamic system, CR acceleration can be 
a very intermittent process;
for instance, a locally enhanced ionisation rate caused by the accelerated particles could modify the local ionisation fraction and then the
efficiency of this acceleration mechanism.
The intrinsic limit of our modelling lies in the observational uncertainties of these parameters. High-resolution observations 
carried out  for example with ALMA will help to have better constraints, with  special consideration for the magnetic field configuration.
In addition, the parameters listed above are not constant all along the shock surface so that the efficiency of the acceleration can be reduced. 
However, while we only accounted for acceleration at the surface of a single inner jet shock at 100~AU 
plus re-acceleration at the final reverse bow shock,  jets usually show
multiple inner shocks, which makes  it possible to recover the acceleration efficiency of our modelling.

After demonstrating the possibility of accelerating CR protons in a jet shock, we described their propagation in the jet, also accounting 
for secondary CR electrons produced by the ionisation process.
Once the reverse bow shock is reached, CR protons and electrons are re-accelerated together with a local component of thermal protons before
streaming in the hot spot region.
A number of observations can be explained by our modelling. In particular:
\begin{itemize}
\item[1.] The synchrotron emission seen in the bow shock of DG~Tau can be explained by the presence of high-energy CR electrons.
Using the available parameter values for DG Tau, we succeeded in reproducing the synchrotron index estimated by observations. The most 
important conclusion is that the electrons responsible for synchrotron emission are the secondary CR electrons re-accelerated at the bow shock.
\item[2.] The high ionisation seen towards the bow shock B1 in L1157 can be reproduced if we assume that the first acceleration of thermal protons
takes place in the shock B0 and that the further acceleration of thermal particles in B1 is inefficient (e.g. because of low 
flow velocity and shock efficiency,
or high upstream diffusion coefficient).
\item[3.] We attempted to describe the high ionisation towards OMC-2 FIR~4. So far no jet activity has been observed, which is why we explained
the observed ionisation rate and  the $^{10}$Be abundance
assuming that CR acceleration takes place directly on the protostellar surface. However, it is important to remember that this source
shows hints of fragmentation so that our modelling may not be directly applied.
 
\end{itemize}
The most limiting condition for the maximum energy reached by a CR proton is due to the geometry of the jet. 
In fact, the jet transverse radius increases with the distance of the shock from the source
and CRs are confined in the jet for a longer time since their possibility of escaping in the perpendicular direction decreases.
In this way they can reach higher energies. Thus, if DSA takes place in a jet shock far from a protostar,
it is possible to accelerate CR protons up to $1-10$~TeV energies and in principle their $\gamma$ emission could be
observed with the next generation of ground-based telescopes such as CTA.

In closing, it is important to note that D'Alessio et al.~(\cite{da98}) found that in the disc region between 0.2~AU and 4~AU the ionisation fraction is smaller than the 
minimum value required to have magnetorotational instability operating near the disc midplane.  
The acceleration of CRs inside protostars predicted by our modelling would cause 
an increase in the number of electrons. As a consequence, the extension of the so called
``dead zone'' (Gammie~\cite{g96}) could be modified together with the planetesimal formation efficiency.

\acknowledgements 
The authors thank Cecilia Ceccarelli, Daniele Galli, and
Vincent Tatischeff 
for valuable discussions.
We acknowledge the financial support of the Agence National
pour la Recherche (ANR) through the COSMIS project. 
This work has been carried out thanks to the support of the OCEVU Labex (ANR-11-LABX-0060) 
and the A*MIDEX project (ANR-11-IDEX-0001-02) funded by the ``Investissements d'Avenir'' French government programme managed by the ANR.
MP and AM also acknowledge the support of the CNRS-INAF PICS project ``Pulsar wind nebulae,
supernova remnants and the origin of cosmic rays''.

\appendix

\section{Alternative acceleration mechanisms}
\label{app:othermechanisms}

\subsection{Shear acceleration at the jet-outflow interface}\label{shear}
At the interface between the jet and the outflow, particles see two fluids moving at different velocities and they can gain energy
by being scattered by magnetic field turbulence in the same manner as in the DSA process except that the up- and downstream media are replaced 
by the outer and inner shear flows.  
In order to establish which of these two mechanisms prevails, we compare the acceleration timescale due to DSA, $t_{\rm acc}$,
given by the inverse of Eq.~\ref{nuacc}, and that due to shear acceleration, $t_{\rm shear}$. 
Following Rieger \& Duffy~(\cite{rd06}), $t_{\rm shear}$ reads
\be
t_{\rm shear}=\frac{1}{4+\alpha}\frac{(\beta c)^{2}}{3\kappa_{\rm shear}\Gamma}\,,
\ee
where $\kappa_{\rm shear}=k_{\rm shear}\kappa_{\rm gal}$, with $k_{\rm shear}=0.01-1$. 
The shear flow coefficient, $\Gamma$, reads (Earl et al.~\cite{ej88})
\be
\Gamma\simeq3\times10^{-18}\left(\frac{v_{\rm jet}}{10^{2}~\mathrm{km~s^{-1}}}\right)^{2}%
\left(\frac{R_{\perp,\,{\rm outflow}}}{10^{2}~\mathrm{AU}}\right)^{-2}~\mathrm{s^{-2}}\,,
\ee
where we assumed $v_{\rm jet}\gg v_{\rm outflow}$ and $R_{\perp,\,{\rm jet}}\ll R_{\perp,\,{\rm outflow}}$.
The parameter $\alpha$ comes from the assumption of momentum-dependent scattering time $\tau\propto p^{\alpha}$
(Rieger \& Duffy~\cite{rd06}).
Using the three different trends for $\kappa_{\rm gal}$ described in Sect.~\ref{twozonemodel},
we find $\alpha=-2$ for a constant diffusion coefficient at non-relativistic energies, $\alpha=-1$ for the DR model, and 
$\alpha=-4$ for the PD model. We note that in this last case, $t_{\rm shear}\rightarrow\infty$ and the shear acceleration does not occur. 
We compute $t_{\rm acc}$ for model $\mathcal{S}$ minimising $t_{\rm shear}$ (by taking $k_{\rm shear}=1$, $v_{\rm jet}=300$~km~s$^{-1}$, and
$R_{\perp,\,{\rm outflow}}=10^{3}$~AU), and we find  
$t_{\rm acc}<t_{\rm shear}$ when $E\lesssim100$~GeV. 
At a greater distance from the protostar, DSA is even more favoured since $t_{\rm shear}\propto R_{\perp,\,{\rm outflow}}^{2}$.
However, in terms of geometrical efficiency, shear 
acceleration could be more efficient than DSA since, in principle, the former process can occur all along the 
jet-outflow contact surface while DSA takes place only at a shock surface. 

\subsection{Acceleration by shocked background turbulence}
The basic requirement for the acceleration mechanism described thus far assumes that the waves providing the scattering of the energetic particles are
self-generated by the particles themselves.
However, turbulence in protostars can be injected by other means since jets propagate into a turbulent environment (Giacalone \& Jokipii 
\cite{gj07}) and because the magneto-rotational instability in the accretion process induces turbulent motions (Hawley \cite{hjf00}). 
The level of background turbulence is poorly known, but it could overcome the turbulent fluctuations generated by any 
CR accelerated at a shock front. 
Finally, it might even be that background turbulence is required to trigger CR acceleration, 
which, if efficient enough, could enter the self-generated regime.

Background turbulence prevails over self-generated turbulence if  
\be
W_{\rm bg} > \frac{P_{p}}{M_{\rm A}}\,,
\ee
where $W_{\rm bg}=\delta B_{\rm bg}^{2}/(8\pi)$ is the background magnetic pressure; $P_{p}\propto p^4 v f(p)$ 
is the particle pressure at momentum $p$, where $p$ is related to the wavenumber $k$ through the resonance condition
$k\propto 1/r_{\rm L}$, $r_{\rm L}$ being the Larmor radius (e.g. Drury et al.~\cite{dd96}); and
\be\label{Malf}
M_{\rm A}=\frac{U}{V_{\rm A}}
\ee
is the
Alfv\'enic Mach number. If a background CR distribution is 
present, then the condition becomes 
\be
W_{\rm bg} > \frac{P_{p}}{M_{\rm A}}\left[1-\frac{f_{\rm bg}(p)}{f(p)}\right]\,.
\ee
If the above condition is fulfilled, DSA theory has to be applied with diffusion coefficients controlled by background turbulence, i.e. diffusion coefficients that deviate from the Bohm scaling given by Eq.~(\ref{diffusioncoeff}).

\subsection{Turbulent second-order Fermi acceleration}
Low-energy particles, in the non- or mildly-relativistic regime, can be subject to stochastic acceleration by the turbulence generated around the shock. The relative importance of stochastic acceleration and pitch-angle scattering is given by the ratio of the two corresponding timescales (Petrosian \& Bykov~\cite{pb08}), namely
\be
\frac{D_{pp}}{p^2D_{\mu\mu}}\approx \frac{V_{\rm A}}{\beta c}^2 \,,
\ee
where $D_{pp}$ and $D_{\mu\mu}$ are the Fokker-Planck coefficients, $\mu$ being the cosine of the pitch angle.
As discussed in Prantzos et al.~(\cite{pb11}), stochastic acceleration 
is important only if it can overcome the particle losses given by Eq.~(\ref{nuloss}).
Second-order Fermi acceleration becomes important when $\beta\sim V_{\rm A}/c$, but for typical ranges of
total hydrogen, ionisation fraction, and magnetic field strength in a protostar (Sect.~\ref{fulfilment}), this acceleration mechanism turns out to be negligible.

\subsection{Magnetic reconnection}
In more evolved young stellar objects, magnetic reconnection is expected in coronal winds above the accretion disc and  in accretion columns from the inner disc region to the central source.
De Gouveia Dal Pino \& Lazarian~(\cite{dl05}) proposed that an acceleration mechanism similar
to DSA takes place in reconnection sites. In fact, particles are expected to gain energy bouncing between converging flux tubes with oppositely directed magnetic fields. Through 3D magnetohydrodynamic numerical simulations, de Gouveia Dal Pino et al.~(\cite{dk14}) found the formation of a hard proton
power spectrum ($\propto E^{-1}$) for proton energies between about 10~GeV and 1 TeV.

\section{Turbulent damping in jets}\label{app:dampingrate}
In a plasma, turbulent perturbations can either be damped by collisional or collisionless processes depending on the ratio between the wavelength of the perturbations to the proton mean free path due to collisions (see Yan \& Lazarian~\cite{yl04}). Here, the resonant waves have a wavelength, $\lambda$, given by
\be
\lambda \simeq r_{\rm L} = 3.3 \times 10^9 \left(\frac{E}{\rm GeV}\right)\left(\frac{B}{\rm mG}\right)^{-1}~\mathrm{cm}\,,
\ee
while the proton mean free path, $\ell_{\rm mfp}$, reads 
\be
\ell_{\rm mfp}=9.4 \times 10^6 \left(\frac{T}{10^{4}~\mathrm{K}}\right)^2\left(\frac{n_{\rm H}}{10^{5}~\mathrm{cm^{-3}}}\right)^{-1}~\mathrm{cm}\,.
\ee
The damping is in the collisional regime if $\lambda > \ell_{\rm mfp}$ otherwise it is in the collisionless regime.
In the conditions prevailing in stellar jets, particles with energies higher than a few MeV generate waves that are damped in the collisional regime. The dominant collisional damping process is due to ion-neutral collisions and the damping rate is given by  Eq. \ref{wavedampingrate}.  
If $E > 1$~GeV, then $\Gamma \approx \omega_n (\omega/\omega_i)^2$, while if \mbox{$E\in[3 \times 10^{-3}, 1]$~GeV}, we have 
$\Gamma \approx \omega_n$.
The damping scale is
\be
L_{\rm d}= U_{\rm d} \Gamma^{-1}= 7.5 \times 10^{10}  \times \left(\frac{U}{100~\mathrm{km~s^{-1}}}\right)%
\left(\frac{\Gamma}{10^{-4}}\right)^{-1}~\mathrm{cm}\,,
\ee
where $U_{\rm d}$ is the downstream flow velocity.
If we assume $\Gamma = \omega_n$, we obtain $L_{\rm d} \sim 4 \times 10^{-4}$ AU, which  is very small with respect to the distance between the internal shock in the jet and the termination shock. 
At high energies, although the damping is weaker and scales as $(E/{\rm GeV})^{-2}$, 
the damping scale is always much smaller than 1~AU, since the typical maximum energy found is of the order of 10~GeV.

At low energies, the damping corresponds to very short wavelengths and occurs in the collisionless regime. The appropriate damping rate is given by  Eq.~(A.13) in Jean et al.~(\cite{jg09}). It corresponds to the linear Landau damping rate, $\Gamma_{\rm LD}$, which is the resonant interaction of the waves with the background thermal plasma. The difficulty here is fixing a value of the wave pitch-angle $\Theta$ (the angle between the wave number and the magnetic field direction). Downstream of the shock, accounting for the magnetic field compression at the shock, this angle is likely large, close to 
90$^{\circ}$, and causes the damping to vanish. However, one can argue that the magnetic field keeps some obliquity downstream. For instance, starting with an upstream pitch angle of 45$^{\circ}$, a compression of the transversal magnetic component produces a downstream pitch angle $\Theta = 63^{\circ}$, 
leading to 
$\Gamma_{\rm LD} \approx 30 F(\Theta) [5.4\times10^{-4} + 0.8G(\Theta)]~\rm{s^{-1}}$, with $F(\Theta) \approx 0.17$ and $G(\Theta) \approx 1.17$, respectively. It follows that the damping length is very small as before, but it increases rapidly for $\Theta \rightarrow 90^{\circ}$ or $0^{\circ}$. Nevertheless, these cases are likely marginal for our modelling and we assume 
that waves are also rapidly damped behind the shock in the collisionless regime. 

\section{Collisionless character of the shocks and thermal equilibration}
\label{app:collisionless}
The collision time between protons and electrons 
and the gyro-period have to be compared
in order to determine whether a shock is collisional or collisionless. 
Equivalently, the proton-electron collision length ($\lambda_{\rm c}$)
can be compared with the Larmor radius ($r_{\rm L,th}$) of a thermal particle.
If $\lambda_{\rm c}>r_{\rm L,th}$, then the shock is collisionless and it is mediated by magnetic field rather than
by collisions. In this case the shock can accelerate thermal plasma particles. Conversely,
if $\lambda_{\rm c}<r_{\rm L,th}$, then the shock is collisional and it can accelerate particles only if these are
already accelerated in the pre-shocked medium and if magnetic fields are also involved. 

The proton-electron collision length reads
\be
\lambda_{\rm c}=v_{\rm th,p}\nu_{\rm c}^{-1}~\mathrm{cm}\,,
\ee
where $v_{\rm th,p}$ is the proton thermal speed and $\nu_{\rm c}$ the proton-electron collision frequency.
These quantities are given by
\be
v_{\rm th,p}=9.09\times10^{5}\left(\frac{T}{10^{4}~\mathrm{K}}\right)^{0.5}~\mathrm{cm~s^{-1}}
\ee
and
\be
\nu_{\rm c,p}=4\times10^{-3}\ln\Lambda\left(\frac{T}{10^{4}~\mathrm{K}}\right)^{-1.5}%
\left(\frac{n_{\rm H}x}{10^{6}~\mathrm{cm^{-3}}}\right)~\mathrm{s^{-1}}\,,
\ee
where $\ln\Lambda\sim20$ is the Coulomb logarithm.
The Larmor radius for a proton reads
\be
r_{\rm L,th}=9.47\times10^{6}\left(\frac{T}{10^{4}~\mathrm{K}}\right)^{0.5}%
\left(\frac{B}{10~\mu\mathrm{G}}\right)^{-1}~\mathrm{cm}\,.
\ee
Then a shock is collisional if
\be
\frac{\lambda_{\rm c}}{r_{\rm L}}=\frac{24}{\ln\Lambda}%
\left(\frac{T}{10^{4}~\mathrm{K}}\right)^{1.5}%
\left(\frac{n_{\rm H}x}{\mathrm{10^{6}~cm^{-3}}}\right)^{-1}%
\left(\frac{B}{\mathrm{10~\mu G}}\right)<1\,.
\label{lambdacoverrl}
\ee
Using the values in Table~\ref{paramspace}, it is possible
to verify that shocks both in accretion flows and jets are collisionless.

Because collisionless shocks are present, we have to justify
the assumption of equal temperature downstream in our calculation ($T_{p,\rm d}=T_{e,\rm d}=T$). 
In fact, since recurrent shocks spaced of about 100~AU are observed along the jets, 
the temperature $T$ entering the equations in Sect.~\ref{csva} is the upstream temperature during the
passage of the first shock, and the downstream temperature for the following shocks.

The temperature of a species $s$ is proportional to its mass and downstream reads
\be
T_{s,\rm d}=\frac{r-1}{r^{2}}\frac{m_{s}U^{2}}{k}\,,
\ee
where $k$ is the Boltzmann constant and $s=p,e$. 
The thermal equilibrium between protons and electrons that have different downstream temperatures is reached after a time $t_{\rm eq}$
given by
\be\label{teq}
t_{\rm eq}=\frac{7\times10^{2}}{\ln\Lambda}\left(\frac{T_{e,\rm d}}{10^{4}~\mathrm{K}}\right)^{1.5}\left(\frac{n_{\rm H}x}{10^{6}~\mathrm{cm^{-3}}}\right)^{-1}~\mathrm{s}\,.
\ee
This is the time needed to balance $T_{e}$ and $T_{p}$ and it has to be compared with the time between the passage of
two consecutive shocks, $\Delta t_{\rm sh}$, which reads
\be\label{Dtsh}
\Delta t_{\rm sh}=1.5\times10^{8}\left(\frac{D_{\rm sh}}{10^{2}~\mathrm{AU}}\right)\left(\frac{U}{10^{2}~\mathrm{km~s^{-1}}}\right)^{-1}~\mathrm{s}\,,
\ee
$D_{\rm sh}$ being the distance between two successive shocks.
If $t_{\rm eq}<\Delta t_{\rm sh}$, then protons and electrons reach the same temperature before another shock arrives.
It is straightforward to verify that the last inequality is always verified for the values in Table~\ref{paramspace}
assuming that shocks in jets are separated by about 100~AU.
This confirms the correctness of our hypothesis, and so  we can keep using the same temperature for protons
and electrons for any shock, the latter being observable.

\section{Evaluation of $E_{\rm damp}$}
\label{app:Ecut}
Equation~(29) in Drury et al.~(\cite{dd96}), assuming the shock velocity to be 
much higher than the Alfv\'en speed, gives 
\be\label{Ecutapp}
2\Gamma\kappa_{\rm B}V_{\rm A}=U^{3}\widetilde{P}_{\rm CR}
.\ee
The wave damping rate, $\Gamma$, reads (Drury et al.~\cite{dd96})
\be\label{wavedampingrate}
\Gamma=\frac{\omega^{2}}{\omega^{2}+\omega_{i}^{2}}\omega_{n}\,,
\ee
with $\omega_{n}=n_{\rm H}(1-x)\langle\sigma v\rangle$ and $\omega_{i}=n_{\rm H}x\langle\sigma v\rangle$.
The average value of the product between the charge exchange cross section and the collision velocity, $\langle\sigma v\rangle$, is given
by Kulsrud \& Cesarsky~(\cite{kc71})
\be\label{sigmav}
\langle\sigma v\rangle\approx8.4\times10^{-9}\left(\frac{T}{10^{4}~\mathrm{K}}\right)^{0.4}~\mathrm{cm^{3}~s^{-1}}\,.
\ee
For resonant waves, $\omega$ is given by the ratio between the Alfv\'en speed and the particle's gyrofrequency
\be\label{resonantwaves}
\omega=7\times10^{-9}\frac{\tilde\mu^{-1}}{\gamma\beta}\left(\frac{B}{10~\mu\mathrm{G}}\right)^{2}\left(\frac{n_{\rm H}}{10^{6}~\mathrm{cm^{-3}}}\right)^{-0.5}%
~\mathrm{s}^{-1}\,.
\ee
Equation~(\ref{Edamp}) is found by substituting Eqs.~(\ref{wavedampingrate})--(\ref{resonantwaves}) in Eq.~(\ref{Ecutapp}).

\section{Justification for the steady-state model}\label{app:lifetime}
In order to justify the steady-state hypothesis in Sect.~\ref{twozonemodel}, we have to verify whether $t_{\rm esc,d}$ (Eq.~\ref{tescd}) 
is lower than the lifetime
$t_{\rm life}$ (Eq.~\ref{tlife}) of a bow shock. 
Here we consider two sources with differing  distances of the reverse bow-shock, $D_{\rm rBS}$: HH~111 and
DG~Tau with $D_{\rm rBS}=7\times10^{4}$~AU and $D_{\rm rBS}=1.8\times10^{3}$~AU, respectively.

\subsection{HH~111}
Morse et al.~(\cite{mh93}) states that the upper limit to the shock velocity at the apex of the bow shock V in HH~111 is $v_{\rm sh}=100$~km~s$^{-1}$,
while the jet velocity is $v_{\rm jet}=400$~km~s$^{-1}$. Then, the shock velocity in the shock reference frame is $U=300$~km~s$^{-1}$ and
$t_{\rm life}\simeq835$~yr.
We assume the hot spot radius to be of the order of the transverse radius of knot V, which is about $2\times10^{3}$~AU (Reipurth et al.~\cite{rh97}), then $t_{\rm adv}$ (Eq.~\ref{tadv}) 
is about
40 yr. Since $t_{\rm diff}\ll t_{\rm adv}$ at any energy,
then $t_{\rm esc,d}\simeq t_{\rm diff}\ll t_{\rm life}$ and we can use the steady-state approximation.

\subsection{DG~Tau}
Eisl\"offel \& Mundt~(\cite{em98}) estimates that knot C, which corresponds to the bow shock, was ejected in 1936, so $t_{\rm life}\simeq80$~yr. In
the same paper they compute $v_{\rm jet}\simeq200$~km~s$^{-1}$. Hartigan et al.~(\cite{hm94}) calculate $v_{\rm sh}\simeq100$~km~s$^{-1}$, then
$U\simeq100$~km~s$^{-1}$. The hot spot radius 
is about 1300~AU (Ainsworth et al.~\cite{as14}) and $t_{\rm adv}\simeq t_{\rm life}$.
For HH~111, $t_{\rm diff}\ll t_{\rm adv}$, the 
steady-state approximation is still  marginally valid for this source.

\end{document}